\documentclass[
superscriptaddress,
final,
reprint,
showkeys,
pre,
aps,
floatfix,
]{revtex4-2}

\usepackage[dvips]{graphicx}
\usepackage[utf8]{inputenc}
\usepackage{amsmath,amssymb,booktabs,latexsym,MnSymbol,bm,makecell}
\usepackage[%
  colorlinks=true,
  urlcolor=blue,
  linkcolor=blue,
  citecolor=blue
]{hyperref}
\usepackage[final]{pdfpages}
\makeatletter
\AtBeginDocument{\let\LS@rot\@undefined}
\makeatother
\begin{document}

\title{Universal productivity patterns in research careers}

\author{Andre S. Sunahara}% ORCID: 0000-0002-4666-9995 %\email{sunaharaseiji@gmail.com}
\affiliation{Departamento de F\'isica, Universidade Estadual de Maring\'a -- Maring\'a, PR 87020-900, Brazil}

\author{Matja{\v z} Perc} %ORCID: 0000-0002-3087-541X %\email{matjaz.perc@gmail.com}
\affiliation{Faculty of Natural Sciences and Mathematics, University of Maribor, Koro{\v s}ka cesta 160, 2000 Maribor, Slovenia}
\affiliation{Department of Medical Research, China Medical University Hospital, China Medical University, Taichung, Taiwan}
\affiliation{Alma Mater Europaea, Slovenska ulica 17, 2000 Maribor, Slovenia}
\affiliation{Complexity Science Hub Vienna, Josefst{\"a}dterstra{\ss}e 39, 1080 Vienna, Austria}
\affiliation{Department of Physics, Kyung Hee University, 26 Kyungheedae-ro, Dongdaemun-gu, Seoul, Republic of Korea}

\author{Haroldo V. Ribeiro}% ORCID: 0000-0002-9532-5195 
\email{hvr@dfi.uem.br}
\affiliation{Departamento de F\'isica, Universidade Estadual de Maring\'a -- Maring\'a, PR 87020-900, Brazil}
\date{\today}

\begin{abstract}
A common expectation is that career productivity peaks rather early and then gradually declines with seniority. But whether this holds true is still an open question. Here we investigate the productivity trajectories of almost 8,500 scientists from over fifty disciplines using methods from time series analysis, dimensionality reduction, and network science, showing that there exist six universal productivity patterns in research. Based on clusters of productivity trajectories and network representations where researchers with similar productivity patterns are connected, we identify constant, u-shaped, decreasing, periodic-like, increasing, and canonical productivity patterns, with the latter two describing almost three-fourths of researchers. In fact, we find that canonical curves are the most prevalent, but contrary to expectations, productivity peaks occur much more frequently around mid-career rather than early. These results outline the boundaries of possible career paths in science and caution against the adoption of stereotypes in tenure and funding decisions.
\end{abstract}
\maketitle

\section*{Introduction}

Scientific productivity is routinely used to measure and assess the performance of researchers, as it quantifies their contributions to the scientific community through scholarly publications~\cite{merton1973sociology}. When combined with other indicators of research quality, productivity plays an important role in determining job placement~\cite{clauset2015systematic}, promotions to tenured positions~\cite{bertsimas2015or}, funding allocation~\cite{stephan2015economics, meirmans2019science}, and in mapping the development of science~\cite{fortunato2018science, liu2023data}. Given its importance, understanding productivity patterns over the course of scientific careers has been a long-standing priority for researchers from various disciplines, and Lehman's monograph is considered a seminal work in this regard~\cite{lehman1953age}. In 1953, he observed that the aggregated contributions of scientists, musical composers, artists, and writers exhibit a pattern of rapid early-career growth followed by a gradual decline in productivity as their careers progressed. This pattern has been consistently observed in various contexts and datasets, and it is often referred to as the ``canonical productivity narrative''~\cite{lehman1953age, lehman1954men, dennis1956age, behymer1975environmental, cole1979age, horner1986relation, levin1991research, simonton1997creative, gingras2008effects, rorstad2015publication, sunahara2021association, spake2022analysis}.

But the notion that there is a universal pattern of productivity across scientific disciplines and demographic groups has been significantly challenged by recent research. Indeed, studies have found evidence for a variety of productivity patterns, including constant~\cite{cole1979age, rorstad2015publication}, decreasing~\cite{over1982does, over1982age, horner1986relation}, increasing~\cite{rorstad2015publication}, and periodic-like~\cite{pelz1966scientists, behymer1975environmental, bayer1977career}. However, many studies have used aggregated data, which may introduce bias due to the ``compositional fallacy''~\cite{simonton1997creative} -- a common issue that arises when trying to infer typical productivity trajectories based on average behavior across many individuals. Other studies have been restricted to a reduced set of career years in specific fields of knowledge~\cite{over1982does, over1982age} and have often relied on linear regression models~\cite{over1982age, over1982does, bayer1977career, levin1991research, rorstad2015publication}, which may not fully capture the complexity of productivity patterns. Some authors have also proposed generative models of productivity curves~\cite{levin1991research, simonton1997creative, rinaldi2000instabilities}, but have been unable to validate these patterns with empirical evidence.

Large-scale studies that investigate individual shapes of productivity trajectories are scarce, with the work of Way \textit{et al.}~\cite{way2017misleading} being one of the few exceptions. Using data from over two thousand computer science faculty members in the U.S. and Canada, they applied a segmented linear model composed of two continuous lines to each researcher's career to evaluate the universality of the canonical productivity narrative. Research has found that almost half of the careers in this dataset is consistent with strictly constant, increasing, or decreasing productivity trajectories. Conversely, only 20\% of the trajectories have been found to exhibit early growth followed by a slow decline in productivity, thus suggesting that the canonical narrative may not be as prevalent as previously thought. However, the use of piecewise regressions limits the emergence of possible nonlinear patterns such as periodic trajectories, and the focus on computer science may limit the generalization of these conclusions to other academic disciplines. Additionally, research so far has ignored that structural changes in the scientific enterprise -- such as the increase in scientific collaboration~\cite{wuchty2007increasing} and pressure to produce in large quantities~\cite{miller2011publish, van2012intended,schimanski2018evaluation} -- may impact the research culture of different cohorts and their productivity trajectories.

Here we investigate the productivity trajectories of over eight thousand scientists from the elite of the Brazilian research community, spanning more than fifty research disciplines. We employ a coherent data-driven approach that combines methods from time series analysis, dimensionality reduction, and network science to cluster productivity trajectories based on their pairwise similarities. Unlike most previous works, our approach considers trajectories individually, accounts for discipline-specific inflation of productivity~\cite{solla1963little, sinatra2016quantifying, sunahara2021association}, the noisy nature of individual productivity trajectories, and possible cohort effects. Importantly, we do not explicitly assume one or a set of predetermined shapes for the productivity curves, which allows us to discern the natural emergence of universal patterns of productivity in scientific careers. Our research identifies productivity patterns that have been only qualitatively hypothesized~\cite{bayer1977career} or found in studies based on aggregated data~\cite{bayer1977career, cole1979age, over1982does, over1982age, horner1986relation, rorstad2015publication}. In particular, we identify six categories of productivity trajectories: constant, u-shaped, decreasing, periodic, increasing, and canonical, with the latter two categories describing almost three-fourths of researchers. Increasing trajectories are much more frequent among early-career researchers than among senior researchers (45\% vs. 19\%), while canonical curves are much more prevalent among senior researchers than among younger scholars (65\% vs. 27\%). However, the initial career years of senior researchers are categorized as increasing trajectories with slightly less prevalence than those found for younger researchers. Only a small fraction of senior researchers with initially increasing productivity trends is able to maintain this pattern, while the majority of the remainder progresses to canonical trajectories. This result highlights the importance of considering cohort and size effects when investigating productivity trajectories, and it indicates that young researchers characterized by increasing trajectories may also progress to canonical patterns in the future.

In what follows, we present these results in detail, and then discuss the influence of funding allocation, tenure positions, and job security in research on the emergence of productivity patterns. We also caution against widely held but fixed and oversimplified assumptions associated with scientific careers, in the hope that the reported universal patterns will open the door for more inclusive and improved evaluation of research productivity. 

\section*{Results}

Our results are based on the academic curricula of 8,493 Brazilian researchers from 56 disciplines who hold the CNPq Research Productivity Fellowship (see Methods for details). This traditional fellowship aims to support the scientific enterprise and has been awarded to scholars producing high-quality research since the 1970s. Scholars holding this fellowship are commonly considered among the elite of Brazilian scientists. All curricula vitae were collected from the Lattes platform (\textit{Plataforma Lattes}, a widely used governmental curriculum platform in Brazil) where CNPq fellows are required to keep their complete and up-to-date records for maintaining or applying for the fellowship. Compared to other databases often used in science of science studies, our dataset has the main advantage of not suffering from author name disambiguation issues as well as it offers a systematic coverage of scientists across the country. We construct raw time series of yearly productivity (number of publications per year) for all researchers, assuming that each career starts after doctorate completion. The researchers in our study have career lengths of at least ten years (the same threshold used by Way \textit{et al.}~\cite{way2017misleading}), and the median career length is 17 years (Fig.~S1~\cite{SI}).

We do not directly use raw productivity series in our analysis. Instead, we take into account three characteristics that may hinder the identification of the most common productivity curves: inflation, different scales, and the noisy nature of productivity series. Scientific productivity has been rising worldwide over the years~\cite{solla1963little, sinatra2016quantifying, sunahara2021association}, and the researchers in our study show an overall increase in productivity of approximately $0.8$ papers per year per decade. This inflation is also discipline-specific (Figs.~S2 and~S3~\cite{SI}). To account for inflation effects, we first deflate the productivity series using the yearly average values of each discipline~\cite{petersen2019methods}, such that the deflated productivity represents the re-scaled number of papers per year as if they were published in 2015. Second, to make productivity trajectories comparable among researchers, we calculate standard score values ($z$-scores) of productivity relative to each researcher from the deflated productivity series. The $z$-scores quantify how many standard deviation units researchers perform above or below their own average productivity and make all time series comparable in scale. Productivity series also have an intrinsic noisy nature that reflects the complex processes involved in producing and publishing scientific papers. The publication year often does not mark the actual completion time of an article, as most papers are not promptly accepted for publication. Thus, lastly, we apply a Gaussian filter to the $z$-scores productivity series to account for these random fluctuations.

\begin{figure*}[!ht]
  \centering
  \includegraphics[width=1\textwidth]{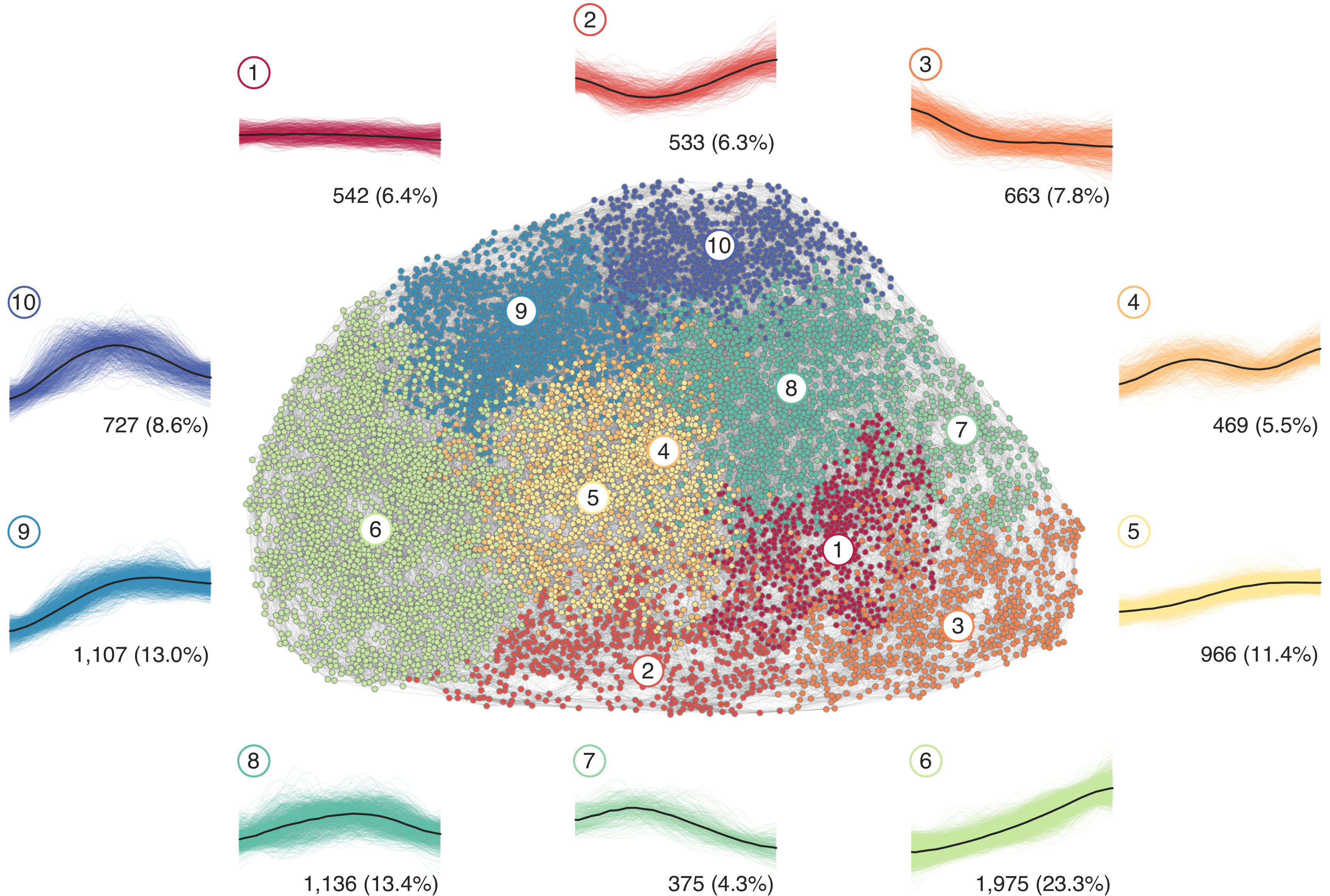}
  \caption{Clustering patterns of researchers' productivity curves. The central panel displays a network representation, where each node represents a researcher and weighted edges connect those with similar productivity trajectories. Ten distinct communities, represented by different colors and labeled 1 to 10, are identified and correspond to groups of researchers with similar productivity patterns. The surrounding panels display the productivity curves of researchers in each community, with the black curves representing the average behavior of each cluster. The lengths of researchers' careers in each group are scaled to the unit interval and the numbers and fractions of researchers in each group are shown within each panel. The ten clusters are further grouped into six categories: constant (cluster 1), u-shaped (cluster 2), decreasing (cluster 3), periodic-like (cluster 4), increasing (clusters 5 and 6), and canonical-like (clusters 7 to 10) curves. Increasing and canonical-like patterns describe almost three-fourths of the researchers in our study, while periodic-like curves are the least common. Clusters and nodes that are close together share similar productivity patterns (see \href{https://complex.pfi.uem.br/cluster}{this page} for an interactive visualization).}
  \label{fig:1}
\end{figure*}

After obtaining deflated, standardized, and smoothed productivity curves, we apply the dynamic time warping (DTW) algorithm~\cite{sakoe1978dynamic} to estimate the similarities among all pairs of researchers' trajectories. The DTW is a shape-based dissimilarity measure that allows the comparison of time series with different lengths and non-optimal alignment -- crucial features for comparing researchers with different career lengths and patterns that can be shifted in time. Next, we use the DTW dissimilarity matrix along with the uniform manifold approximation and projection (UMAP) method~\cite{mcinnes2018umap} to create a network representation of the similarities among researchers' trajectories. UMAP is a state-of-the-art dimensionality reduction technique based on the mathematical grounds of Riemannian geometry and algebraic topology capable of balancing the emphasis between local and global structures~\cite{mcinnes2018umap}. In short, it creates a graph representation from a dissimilarity matrix of high-dimensional datapoints and projects them into a lower-dimensional space using a force-directed layout algorithm. We focus only on the first step of the UMAP algorithm, mapping our dissimilarity matrix into a network where researchers are represented as nodes and weighted edges connect researchers with similar productivity trajectories. Finally, we apply the Infomap algorithm to identify the community structure of the network created by UMAP, which corresponds to groups of researchers with similar productivity trajectories. A similar approach has been recently and successfully used by Lee \textit{et al.}~\cite{lee2021non} to cluster extracellular spike waveforms in a Neuroscience context. All steps used to cluster productivity trajectories are further detailed in the Methods section and illustrated in Fig.~S4~\cite{SI}.

While the final low-dimensional embeddings produced by UMAP are not deterministic (meaning that UMAP yields similar but different embeddings), the network created in its first step is always the same for a fixed dataset. However, the Infomap algorithm is based on probability flows of random walks on the network and produces similar but different network partitions. To account for this non-deterministic nature, we run one thousand realizations of the Infomap algorithm and observe that all partitions are qualitatively comparable. The number of detected communities ranges from 7 to 14, but almost 85\% of all realizations yield from 9 to 11 communities, with 10 being the most common number of partitions (34\%, Fig.~S5~\cite{SI}). We select the best partition as the one with the largest silhouette score~\cite{rousseeuw1987silhouettes} among all realizations with 10 communities (see Methods for details). We use Infomap because it is one of the best-performing methods for detecting planted partitions in benchmark graphs~\cite{lancichinetti2009community, fortunato2010community, fortunato2016community}, particularly for undirected and weighted networks as in our case. However, deterministic community detection methods such as the Louvain~\cite{blondel2008fast} and the Leiden~\cite{traag2019louvain} also generate similar clustering patterns (Figs.~S6 and S7~\cite{SI}), but with lower silhouette scores (Fig.~S8~\cite{SI}).

The central panel of Figure~\ref{fig:1} displays the network representation produced by UMAP, with different colors indicating the ten communities detected by Infomap's best partition. Surrounding the network visualization, we plot the productivity trajectories of all researchers in each group, as well as the average behavior of each cluster (labeled 1 to 10). We also re-scale the lengths of researchers' careers in each group to the unit interval to better visualize trajectories with different lengths. Productivity trajectories in each group display very similar shapes and the silhouette score of the clustering is significantly higher than values obtained by shuffling trajectories among clusters (Fig.~S8~\cite{SI}). Our best partition not only generates internally consistent groups, but it also yields a significantly higher silhouette score compared to null models in which artificial careers are generated from a binomial distribution and the shuffling of productivity trajectories of each researcher (Fig.~S8~\cite{SI}). This network representation preserves both local and global structures of the dissimilarity matrix (Fig.~S9~\cite{SI}), meaning that nodes and clusters that are close together share similar productivity patterns. For example, clusters 7-10 all have an average behavior marked by a peak in productivity and appear adjacent to each other in the network. In contrast, clusters 3 and 6 represent opposite behaviors (increasing vs. decreasing trends) and are therefore located far apart. When visually inspecting productivity patterns over the network representation (see \href{https://complex.pfi.uem.br/cluster}{this page} for an interactive visualization), we also observe that nodes located close to the frontiers between two or more communities often display more complex productivity patterns that may resemble a mixture of the average behavior of adjacent clusters.

Our analysis uncovers a diverse set of productivity trajectories that go beyond the canonical narrative and include patterns that were only conjectured or observed in studies using aggregated data~\cite{bayer1977career, cole1979age, over1982does, over1982age, horner1986relation, rorstad2015publication}. A detailed examination of the trajectories and their derivatives (Fig.~S10~\cite{SI}) allows us to group the ten clusters into six categories: constant (cluster 1), u-shaped (cluster 2), decreasing (cluster 3), periodic-like (cluster 4), increasing (clusters 5 and 6), and canonical-like (clusters 7 to 10) curves. Constant trajectories, which make up 6.4\% of researchers, are characterized by stable or slightly decreasing productivity. U-shaped trajectories, accounting for 6.3\% of researchers, show a decline before an increase in productivity. Decreasing trajectories, representing 7.8\% of researchers, exhibit a sharp decline in the first half of careers followed by an almost constant plateau in productivity. Periodic-like trajectories, which constitute 5.5\% of researchers, have a peak before mid-career followed by a decline before another increase in productivity. Together, these patterns represent slightly more than a quarter of researchers, with periodic-like patterns being the least common. As a result, increasing and canonical-like patterns describe almost three-fourths of the researchers in our study. Specifically, 35\% of researchers display increasing curves, which are divided into two clusters: one where productivity always increases over careers (cluster 6) and the other exhibiting growing trends with declining rates or approaching a plateau (cluster 5). Canonical-like curves, broadly defined here as careers containing a single peak in productivity (clusters 7 to 10), are the most frequent type of trajectory, comprising 39\% of researchers in our dataset. We use the term canonical-like because Lehman's definition is more restrictive, assuming the canonical narrative as ``curves of creativity that rise rapidly in early maturity and then decline slowly after attaining an earlier maximum''~\cite{lehman1953age}. Although this definition is qualitative, one may interpret that solely cluster 7 strictly meets Lehman's definition, as it is the only cluster that shows a maximum before mid-career (Fig.~S11~\cite{SI}). The peak positions are indeed one of the most distinct behaviors among clusters 7 to 10, and the reason they emerge as separated clusters (Table~S1~\cite{SI}).

To validate the robustness of the six categories of productivity trajectories, we perform ten realizations of our clustering procedure using subsamples obtained by randomly dividing our dataset into three equal-sized parts. For every part and realization, we verify that the clusters can be categorized into the same six patterns observed in the complete data. We classify each researcher into one of the six categories in each realization, allowing us to verify the consistency with the classification obtained from the entire dataset. On average, 73\% of researchers are assigned to the same category as determined from the full data. The confusion matrix primarily exhibits a diagonal pattern, with inconsistencies occurring mainly when periodic curves are labeled as increasing or canonical trajectories (Fig.~S12A~\cite{SI}). We also calculate the normalized entropy related to the assignment probabilities of each pattern for every researcher across the ten realizations. This analysis reveals that 80\% of researchers display normalized entropy below 0.5, indicating low variability in their assigned category (Fig.~S12B~\cite{SI}). Moreover, approximately one-third of researchers exhibit zero entropy, signifying that they are consistently assigned to a single category. We further observe that researchers displaying higher entropy are located in the frontier between two or more clusters (where patterns tend to be more complex) as well as in the region of overlap between the periodic (cluster 4) and increasing with declining rates (cluster 5) trajectories in the network representation (Fig.~S12C~\cite{SI}). These same observations hold true when dividing the dataset into two halves (Fig.~S13~\cite{SI}). 

Additionally, we conduct a human validation where a panel of two experts categorizes 25\% of trajectories randomly sampled from our dataset in a stratified manner. They are introduced to an interactive application where $z$-scores and smoothed trajectories are individually shown. Buttons are provided for each category, and an additional button is available when they disagree on the classification, for performing the task. We compare these human-based labels with those determined from our clustering procedure, finding an overall agreement of 73\% and a confusion matrix mostly diagonal (Fig.~S14A~\cite{SI}). Inconsistencies occur primarily when experts classify decreasing trajectories as u-shaped curves and periodic trajectories as canonical curves. Periodic and u-shaped curves are also the categories with the highest levels of disagreement between the experts. Among increasing and canonical categories, the increasing with declining rates (cluster 5) and late peak (cluster 9) productivity curves are most frequently confused with each other (Fig.~S14B~\cite{SI}). Similar to the subsampling validation analysis, disagreements between the experts' classification and our clustering process occur for careers located in the frontier between two or more clusters and in the overlapping region between periodic (cluster 4) and increasing with declining rates (cluster 5) patterns (Fig.~S14C~\cite{SI}).

\begin{figure*}
  \centering
  \includegraphics[width=1\textwidth]{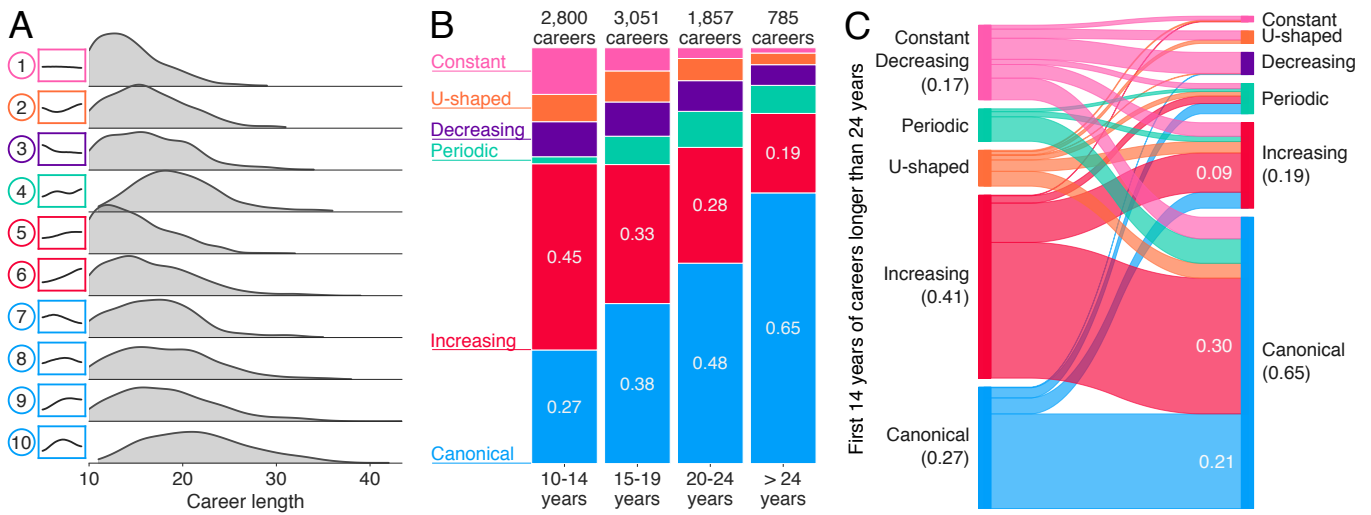}
  \caption{Career length and cohort effects on the prevalence of productivity patterns. (A) Probability distributions of career lengths for each of the ten clusters of productivity trajectories, as determined by kernel density estimation. All clusters encompass a broad range of career lengths, but these distributions are more localized in distinct positions (Table~S1~\cite{SI}). (B) Prevalence of productivity patterns across four categories of career length: 10-14 years, 15-19 years, 20-24 years, and greater than 24 years. The dominant pattern among researchers with shorter careers, which also correspond to younger scholars, is the increasing productivity curve. This pattern becomes less prevalent among researchers with longer careers, which corresponds to more experienced scholars. Canonical-like trajectories exhibit the opposite behavior and are significantly more prevalent among senior researchers. Periodic-like curves are also more common among researchers with long careers, while constant, u-shaped and decreasing trajectories occur more among young researchers. (C) Comparison of the prevalence of productivity patterns in the initial career years of senior researchers with those exhibited in later career stages. The left bars show the fractions of each productivity pattern obtained when considering the initial 14 career years of researchers with careers longer than 24 years, and the right ones show the prevalence of patterns when considering the full range of their careers. The connections between the left and right bars indicate the migration flow among the productivity patterns. Almost half of canonical senior careers are classified as increasing curves in their beginnings; however, only 9\% of senior researchers who exhibit early-career increasing productivity sustain this pattern with career progression.
  }
  \label{fig:2}
\end{figure*}

The prevalence of each productivity pattern may vary among academic careers with different lengths. To examine this potential size effect, we estimate the career size distributions of researchers in each cluster. Figure~\ref{fig:2}A shows that all clusters encompass a broad range of career lengths, but with distinct median career sizes (Table~S1~\cite{SI}). Constant and increasing curves exhibit the smallest median career sizes (median of $\sim$15 years), while canonical and periodic-like trajectories represent more senior researchers (median of $\sim$20 years). To identify the most common productivity pattern at each career stage, we group academic careers into four length categories (10-14, 15-19, 20-24, and larger than 24 years) and calculate the prevalence of each pattern. Figure~\ref{fig:2}B shows that increasing trajectories are the dominant pattern for short careers, accounting for 45\% of researchers in the shortest career category. However, increasing curves become less prevalent among researchers with longer careers, representing only 19\% of researchers in the longest career category. Canonical-like trajectories present the opposite behavior and are much more prevalent among researchers with longer careers. Only 27\% of the researchers with 10-14 career years display canonical-like productivity trajectories, whereas this pattern characterizes 65\% of researchers with more than 24 career years. Even when combined, constant, u-shaped, decreasing, and periodic-like careers occur less frequently than increasing or canonical-like curves in all length categories. Still, we observe that constant, u-shaped, and decreasing trajectories are relatively more common among younger researchers, while periodic-like curves appear more often among researchers with careers longer than 14 years.

Overall, we find similar occupation trends when analyzing the individual behavior of clusters comprising increasing and canonical-like curves (Fig.~S15~\cite{SI}). However, some clusters are more prevalent across the length categories. The always-increasing pattern (cluster 6) is more frequent than the increasing with declining rate pattern (cluster 5) in all length categories, but especially among the most experienced researchers. Almost all researchers exhibiting increasing trajectories with careers longer than 24 years belong to cluster 6. Among the canonical-like curves, the middle and later-career peak patterns of clusters 8 and 9 are the most common behaviors across all length categories, except among the most experienced researchers, for which cluster 10 is the most common. The early-stage peak behavior of cluster 8 is the rarest pattern across all categories, except for the youngest researchers, and it is the only canonical-like curve whose prevalence does not increase with career length. We also obtain similar occupation trends when considering disciplines separately, with only Biochemistry exhibiting an almost constant fraction of increasing productivity curves across the categories of career length. There are however appreciable differences in the prevalence of specific patterns among disciplines (as detailed in Figs.~S16 and S17~\cite{SI}). For example, canonical-like curves are 8.9 times more frequent than increasing curves among the most experienced mathematicians and only 1.4 times more prevalent among senior biochemists. Conversely, increasing curves are twice as common as canonical-like ones among the youngest mathematicians and chemists and only 1.3 times more prevalent among physicists.

Career length is directly linked to the year of doctorate completion of each researcher (Fig.~S18~\cite{SI}) and serves as a proxy for grouping different generations of scientists. Indeed, the overwhelming majority of researchers with 10-14 career years concluded their doctorates after the 2000s, while those with more than 24 career years did it before the 1990s. These groups of young and senior scientists represent unique cohorts that are subject to specific socioeconomic conditions, cultural environments, knowledge base of the field, and baseline level of research ability~\cite{bayer1977career, levin1991research}. Therefore, the different prevalence of productivity curves may partially reflect the distinct research and publication cultures of these groups. In particular, the much higher fraction of increasing trajectories among the youngest cohort seems to align with the increasing pressure on scholars to produce in large quantities~\cite{miller2011publish, van2012intended,schimanski2018evaluation} and with the fact this pressure is considered exceptionally high on young scientists~\cite{powell2016junior}. At the same time, the careers of young researchers cannot be regarded as complete careers as even patterns emerging after 10-14 career years may change over time. For instance, part of the increasing patterns exhibited by young researchers may eventually represent only the beginning of canonical-like productivity curves. The precise identification of generational effects in the prevalence of productivity patterns thus requires a dataset comprising entire careers of different scientist cohorts, which is not the case in our study.

However, we can partially test this hypothesis by analyzing the initial career years of senior scientists and comparing the prevalence of productivity patterns with the youngest cohort. To do this, we apply our clustering approach to the entire dataset, but only consider the initial 14 career years of researchers with careers longer than 24 years. The best Infomap partition is again formed by ten clusters (Figs.~S19 and~S20~\cite{SI}) with average patterns very similar to those reported in Figure~\ref{fig:1}. This allows us to group them into the same six categories, with only the constant and decreasing patterns (clusters 1 and 2) merged into a single cluster (cluster 1 of Fig.~S19~\cite{SI}). Figure~\ref{fig:2}C shows the prevalence of productivity patterns associated with the beginning of senior careers and the patterns they evolve to when considering the entire length of senior careers (Fig.~S21~\cite{SI} details the transitions among individual clusters). Corroborating our hypothesis, we observe that almost half of the senior careers classified as canonical are classified as increasing curves in their beginnings. Only 9\% of senior researchers exhibiting early-career increasing productivity sustain this pattern with career progression. Conversely, 78\% of senior researchers with canonical-like early careers maintain this pattern in later career stages. Moreover, about 21\% of senior careers classified as canonical show an initial part compatible with constant/decreasing, periodic, and u-shaped patterns. These rarer transitions are usually associated with careers localized in the border between two or more communities, representing thus more complex productivity patterns (see Fig.~S22~\cite{SI} for examples).

The behavior of senior scientists may not predict the future for young scholars, but our findings suggest that the high prevalence of increasing productivity patterns among young researchers reflects the incomplete nature of their careers. If early-career researchers follow their senior counterparts, much more researchers will likely have productivity patterns represented by canonical curves in the future. However, we cannot ignore the potential effects of generational differences when comparing the productivity patterns of young researchers even with the initial career years of senior researchers. Indeed, our results show that increasing patterns are 10\% more common among young researchers, while periodic-like curves are three times more frequent in the initial years of senior careers (Figs.~\ref{fig:2}B and \ref{fig:2}C). At the same time, these early differences are relatively small, suggesting that the structural changes in the scientific enterprise~\cite{moher2018assessing, schimanski2018evaluation, meirmans2019science} may have only a minor impact on researchers' productivity trajectories.

\section*{Discussion and Conclusions}

We have performed a comprehensive analysis of productivity trajectories for over eight thousand researchers from 56 different research disciplines. Unlike previous studies that have focused on specific disciplines~\cite{over1982age, over1982does, way2017misleading, spake2022analysis}, inferred typical productivity curves from averaged behavior~\cite{lehman1953age, lehman1954men, dennis1956age, pelz1966scientists, behymer1975environmental, bayer1977career, cole1979age, over1982age, over1982does, horner1986relation, levin1991research, gingras2008effects, rorstad2015publication, spake2022analysis}, or assumed particular forms of productivity trajectories beforehand~\cite{levin1991research, simonton1997creative, rinaldi2000instabilities, way2017misleading}, we have evaluated pairwise similarities among these trajectories, and accounted for inflation, different scales, and random fluctuations of productivity curves. Moreover, our research uses a comprehensive dataset with no issues involving name disambiguation that offers systematic coverage of Brazilian scientists across different areas and generations, which in turn contributes to reducing the so-called ``WEIRD bias''~\cite{arnett2016neglected} in science of science studies. Our approach revealed clusters of productivity trajectories that are internally consistent, more cohesive than null models, and robust against data subsampling, as well as that are in semantic agreement with human validation. In addition, our clustering procedure resulted in a network representation where researchers and clusters with similar productivity patterns are closely connected. We have uncovered a range of productivity patterns that go beyond the traditional narrative and can be classified into six universal categories: constant, u-shaped, decreasing, periodic-like, increasing, and canonical-like curves. When combined, constant, u-shaped, decreasing, and periodic-like curves account for slightly more than a quarter of researchers, while the majority of researchers, nearly three-fourths, exhibit canonical-like or increasing patterns. 

We have also investigated possible career length and cohort effects on the prevalence of the different productivity patterns. This analysis has revealed that all clusters encompass a broad range of career lengths, but increasing productivity curves are the dominant pattern among researchers with shorter careers, who are also younger scholars, while canonical-like curves are the most common pattern among senior researchers. We have hypothesized that the higher incidence of increasing productivity patterns among younger scholars may be linked to changes in the scientific enterprise, such as increased collaboration~\cite{wuchty2007increasing, danus2023differences} and pressure on scholars (particularly on young scientists~\cite{powell2016junior}) to publish in large quantities~\cite{moher2018assessing, schimanski2018evaluation, meirmans2019science}, but also to the fact that early-career patterns may evolve as young researchers progress in their careers. While identifying clear generational effects in the prevalence of productivity patterns would require data on the entire careers of different scientist cohorts, we have partially tested our hypotheses by comparing the initial career years of senior scientists with the careers of young scholars. These results showed that almost half of the canonical-like curves among senior researchers are classified as increasing patterns in the beginning. Conversely, only 9\% of senior researchers who exhibited early-career increasing productivity sustained this pattern as their careers progressed. The relatively small differences in the prevalence of patterns observed between young researchers and the initial career years of senior researchers suggest that the behavior observed for senior scientists does not necessarily dictate the career trajectory of young scholars. However, if early-career researchers follow the same trajectory as their senior counterparts, the prevalence of canonical-like curves is likely to be underestimated.

But even if possibly underestimated, canonical-like curves -- broadly defined here as careers with a single peak in productivity -- are the most prevalent productivity pattern, accounting for almost two-fifths of researchers. While this result somehow supports the canonical narrative of scientific productivity, we have also observed that less than 5\% of researchers in our study strictly meet Lehman's ``canonical productivity narrative''~\cite{lehman1953age} and exhibit productivity curves that ``rise rapidly in early maturity and then decline slowly after attaining an earlier maximum''~\cite{lehman1953age}. These researchers belong to cluster 7, which is only one of four clusters that is classified as canonical-like, have median career lengths of 17 years, and present a peak in productivity approximately 6 years after their doctorates. The other three clusters (8, 9, and 10) account for almost 90\% of researchers with canonical-like patterns, who have slightly larger median career lengths but a peak in productivity around 12 years after their doctorates. Although the ``earlier maximum'' in Lehman's definition is subjective, our research shows that the peak in productivity is more likely to occur around mid-career rather than early-career. Additionally, the rise and decline in the productivity of researchers observed in our study is much more varied than in Lehman's definition.

We have further revealed that, when focusing on the initial career years, most researchers in our study exhibit an increasing productivity pattern. This initial trend emerges among clusters 4 to 10 and accounts for approximately 80\% of them. The high incidence of increasing productivity patterns in early-career stages can likely be attributed to the way funding and hiring decisions are made in academia. Research has shown that productivity plays a significant role in determining job placement~\cite{clauset2015systematic} and access to financial resources needed to continue research~\cite{nuffield, dora, hicks2015bibliometrics, stephan2015economics, wilsdon2016metric, meirmans2019science}. Therefore, it is likely that the prevalence of early-rising trends in productivity reflects the tendency to reward more productive researchers. On the other hand, about half of the researchers in our sample (those belonging to clusters 7 to 10 and 3) exhibit a decline in productivity that is more often observed after mid-career stages. Several hypotheses may account for this pattern. For example, the consolidation of academic prestige in late-career stages may reduce the urgency of maintaining high productivity~\cite{bourdieu1975specificity}. The tension between time spent performing scientific research, which is arguably often larger for young researchers, and administrative tasks, which in turn is usually larger for senior researchers, may also be partly responsible for the decline in productivity during late-career stages~\cite{bourdieu1991peculiar, bourdieu2004science, rorstad2015publication}. Parenthood may also contribute to a drop in productivity since time spent on research is typically reduced in such circumstances~\cite{morgan2021unequal}. Finally, the hardly avoidable decline in intellectual potential over time may also be related to a reduction in productivity with career progression~\cite{simonton1997creative}. 

In conclusion, our research reveals that the scientific productivity of a significant number of researchers increases during their early careers and declines after reaching mid-career. However, the presence of six universal productivity patterns and the wide variability among different cohorts caution against relying on stereotypes in funding and tenure decisions. We hope that our findings will inspire further investigations into the characteristics that define each cluster of researchers and contribute to a more comprehensive and inclusive evaluation of scholarly performance. 

\section*{Methods}

\subsection*{Data}\label{sec:data}

The dataset used in our study was extracted from the Lattes Platform (\textit{Plataforma Lattes})~\cite{lattes}. This platform is hosted and maintained by the Brazilian National Council for Scientific and Technological Development (CNPq -- \textit{Conselho Nacional de Desenvolvimento Cient\'ifico e Tecnol\'ogico}), a governmental agency that promotes scientific and technological research in Brazil. The Lattes Platform contains a consolidated national database of curriculum vitae (CV), research groups, and institutions in a standardized form. Furthermore, the Lattes CV has become the official curriculum vitae for Brazilian researchers and is widely used by science funding agencies and universities in performance evaluations. The platform contains a wide range of information for each researcher, including basic data such as discipline, workplace history, and current affiliation, as well as more detailed information such as academic mentorship relationships and scientific production records. Compared to other datasets, our data based on the Lattes Platform has the main advantage of solving issues related to author name disambiguation as well as ensuring comprehensive coverage of scientists across diverse academic disciplines.

We initially selected the CVs of the 14,487 researchers from 88 disciplines holding the CNPq Research Productivity Fellowship as of May 2017. The total scientific output of these researchers comprises 1,121,652 publications. The CNPq fellowship has been awarded to scholars presenting outstanding scientific impact and innovation in their respective areas of knowledge since the 1970s. These researchers, commonly regarded as the elite of the Brazilian scientific community, are required to maintain a complete and up-to-date record of their research activities on the Lattes Platform. To construct the productivity trajectories, we collected the yearly publication records of each researcher starting from the doctorate completion date. We filled in missing information using the CrossRef API (via the DOI reference of the papers) and filtered out researchers with missing doctorate conclusion date or discipline information.  Additionally, we only considered researchers with ten or more career years, the same threshold used by Way \textit{et al.}~\cite{way2017misleading}. 

\subsection*{Deflated, standardized, and smoothed productivity series}

The volume of scientific production has been consistently increasing over time, as observed in both individual and aggregate productivity levels~\cite{solla1963little, sinatra2016quantifying, sunahara2021association}. However, this increase in productivity, or productivity inflation, does not affect all disciplines equally and is likely influenced by varying publication practices among them~\cite{dundar1998determinants, foster2015tradition, sunahara2021association}. In our study, we found that researchers present an overall rise in productivity of approximately 0.8 papers/year per decade, and this inflation varies among disciplines (Figs.~S2 and~S3~\cite{SI}). For example, while productivity has increased by approximately 2.1 papers/year per decade among researchers working in medicine, it has only risen by approximately 0.7 papers/year per decade among physicists. To account for this discipline-specific inflation, we followed Petersen \textit{et al.}~\cite{petersen2019methods} and calculated a deflated measure of productivity defined as
\begin{equation*}
   p_j(y) = \overline{p}_j(y) \dfrac{\mu_{p}(2015)}{\mu_{p}(y)}\ ,
\end{equation*}
where $\overline{p}_j(y)$ is the raw productivity of researcher $j$ in year $y$ and $\mu_{p}(y)$ is the average value of productivity of his/her discipline in year $y$. We used the Huber robust estimator~\cite{huber2004robust} for location (as implemented in the Python package \textit{statsmodels}~\cite{seabold2010statsmodels}) to estimate the average productivity of each discipline and account for outlier observations (Fig.~S23~\cite{SI}). Additionally, we only estimated the average productivity of disciplines for years containing the publication of records of at least 50 researchers, discarding all researchers with at least one year without their discipline's average productivity estimate. This approach yielded our final dataset comprising the deflated productivity trajectories of 8,493 researchers divided into 56 research disciplines (Fig.~S24~\cite{SI}).

To make the deflated productivity trajectories comparable in scale, we further standardized their values by calculating the $z$-score productivity $P_j(y)$ for researcher $j$ in year $y$ via
\begin{equation*}
   P_j(y) = \dfrac{p_j(y) - \mathbb{E}[p_j]}{\mathbb{S}[p_j]}\ ,
\end{equation*}
where $\mathbb{E}[p_j]$ is the average and $\mathbb{S}[p_j]$ is the standard deviation of deflated productivity along the entire career of researcher $j$. The noisy nature of productivity trajectories (Fig.~S25~\cite{SI}) also poses a challenge in estimating dissimilarity measures among them. These fluctuations reflect the intrinsic nature of scientific publishing, as every work goes through a time-consuming and non-deterministic process of reasoning, testing, writing, and peer-review evaluation~\cite{bjork2013publishing}. The exact point in time a paper is published often does not reflect the actual completion time of the work. To address this issue, we applied a Gaussian smoothing filter to all $z$-score productivity series (as implemented in the Python \textit{SciPy} package~\cite{virtanen2020scipy}). This filter assigns Gaussian weights with a standard deviation $\sigma$ centered on each data point and uses these weights to average the time series values through a convolution process. The parameter $\sigma$ controls the degree of smoothing and defines a time scale for averaging the productivity values over neighboring years (Fig.~S26~\cite{SI}). We used $\sigma=2$ years for all results in the main text, but similar clustering patterns were obtained when varying $\sigma$ from $1.0$ to $2.5$ years in half-year intervals (Figs.~S27, S28, and S29~\cite{SI}). By applying the Gaussian filter on the $z$-score productivity trajectories, we ensured that the smoothing was uniformly applied across researchers with different productivity variability.

\subsection*{Time series clustering}

We estimated the similarities between all pairs of pre-processed productivity trajectories using the dynamic time warping (DTW) algorithm~\cite{sakoe1978dynamic} (as implemented in the Python package \textit{dtaidistance}~\cite{dtai}). The DTW is a shape-based dissimilarity measure that allows for optimal alignment of sequences by creating a non-linear warping path between them, providing more flexibility for matching sequences that show similar patterns but are shifted in time. The resulting dissimilarity matrix was then used as a precomputed metric in the uniform manifold approximation and projection (UMAP) dimensionality reduction algorithm~\cite{mcinnes2018umap} (as implemented in the Python package \textit{umap}~\cite{mcinnes2018umapsoftware} and with default parameters). UMAP first creates a fuzzy simplicial complex, which can be represented as a weighted graph, and then projects the data into a lower-dimensional space via a force-directed graph layout algorithm. The first step of the algorithm thus creates a network representation of the dissimilarity matrix, where nodes represent researchers and weighted edges connect researchers with similar productivity trajectories. 

Following the recent work of Lee \textit{et al.}~\cite{lee2021non} in Neuroscience, we used only the network topological structure and discarded the low-dimensional representation produced by UMAP, mapping thus the clustering of time series into a community detection problem. Specifically, we applied the map equation~\cite{rosvall2008maps, rosvall2009map} and the hierarchical map equation~\cite{rosvall2011multilevel}, the so-called Infomap approach, to determine the community structure of the UMAP network. Infomap is a network clustering technique based on concepts of information theory that relies on random walks as a proxy for information flow over the network. This method is one of the best-performing in detecting planted partitions in benchmark graphs~\cite{lancichinetti2009community, fortunato2010community, fortunato2016community} and is capable of identifying network partitions (clusters and sub-clusters) where the random walker is more likely to spend time. The map equation and the hierarchical map equation represent the theoretical limits of how concisely one can describe an infinite random walk on the network (the description length) with a particular partition configuration. By minimizing the map equation or the hierarchical map equation, Infomap uncovers the community structure of the network. We used the Infomap implementation available in the Python package \textit{infomap}~\cite{infomap} with default parameters and tested both the standard two-level model and the hierarchical model. We verified that the hierarchical map equation more effectively estimates the network's modular structure (that is, it yields smaller description lengths when compared with the two-level model) and was therefore chosen as our clustering algorithm. We ran one thousand realizations of the Infomap algorithm by varying the seed parameter in each iteration and found visually similar community structures. However, we defined the best network partition as the one maximizing the silhouette coefficient~\cite{rousseeuw1987silhouettes} among all partitions with the modal number of clusters. Finally, we verified that the Louvain~\cite{blondel2008fast} and the Leiden~\cite{traag2019louvain} (as used by Lee \textit{et al.}~\cite{lee2021non}) community detection algorithms also resulted in similar clustering patterns (Figs.~S6 and S7~\cite{SI}).

\begin{acknowledgments}
The authors acknowledge the support of the Coordena\c{c}\~ao de Aperfei\c{c}oamento de Pessoal de N\'ivel Superior (CAPES), the Conselho Nacional de Desenvolvimento Cient\'ifico e Tecnol\'ogico (CNPq -- Grant 303533/2021-8), and the Slovenian Research Agency (Grants J1-2457 and P1-0403). 
\end{acknowledgments}

\bibliography{references}

\clearpage
% \clearpage


\section*{Supplementary material}

# Universal productivity patterns in research careers

Andre S. Sunahara,[1] Matjaž Perc,[2, 3, 4, 5, 6] and Haroldo V. Ribeiro[1, *]

[1]*Departamento de Física, Universidade Estadual de Maringá – Maringá, PR 87020-900, Brazil*
[2]*Faculty of Natural Sciences and Mathematics, University of Maribor, Koroška cesta 160, 2000 Maribor, Slovenia*
[3]*Department of Medical Research, China Medical University Hospital, China Medical University, Taichung, Taiwan*
[4]*Alma Mater Europaea, Slovenska ulica 17, 2000 Maribor, Slovenia*
[5]*Complexity Science Hub Vienna, Josefstädterstraße 39, 1080 Vienna, Austria*
[6]*Department of Physics, Kyung Hee University, 26 Kyungheedae-ro, Dongdaemun-gu, Seoul, Republic of Korea*
(Dated: October 27, 2023)

## SUPPLEMENTARY INFORMATION

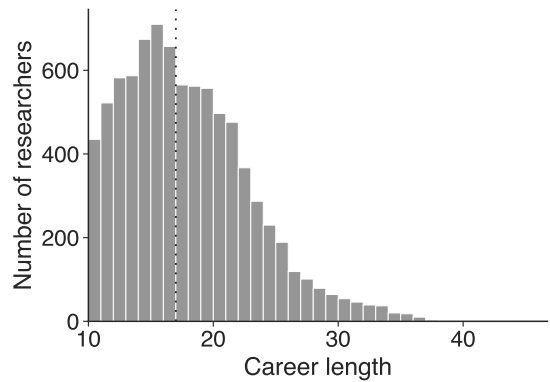

Figure S1. Distribution of career lengths of researchers in our study. The bars represent a histogram of the career lengths, with the vertical dotted line indicating the median career length. Researchers' careers begin after the completion of their doctorate and all researchers in our study have a career length of at least ten years.

* hvr@dfi.uem.br

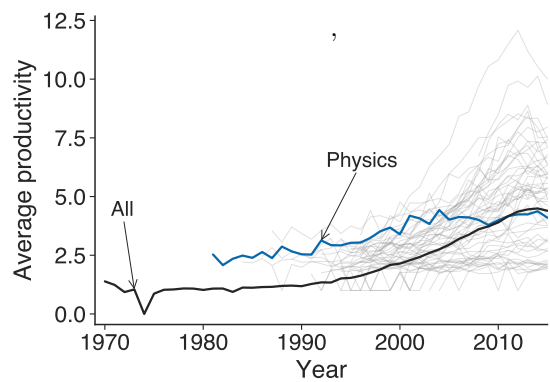

Figure S2. The increase in productivity, or productivity inflation, across different disciplines in our study. The gray curves represent the average productivity of each discipline, with the blue curve highlighting the trend for Physics. The black curve represents the overall trend when aggregating productivity across all disciplines.

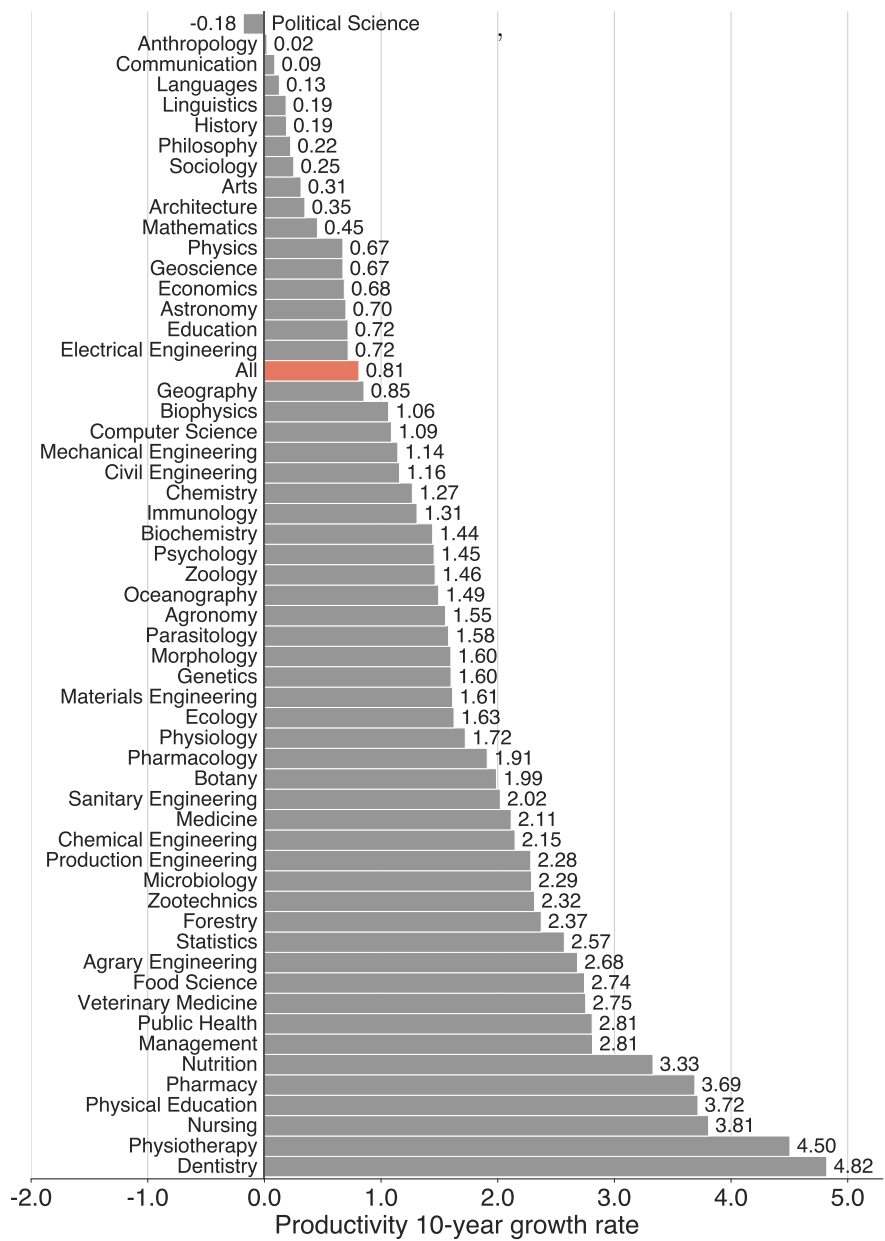

Figure S3. Discipline-specific productivity growth rates. The bar plot shows the per-decade productivity growth rate for each discipline in our study, as well as for the overall behavior of all disciplines combined (red bar). These rates were calculated by fitting a linear model to the evolution of the average productivity for each discipline and to the aggregate behavior of all of them.

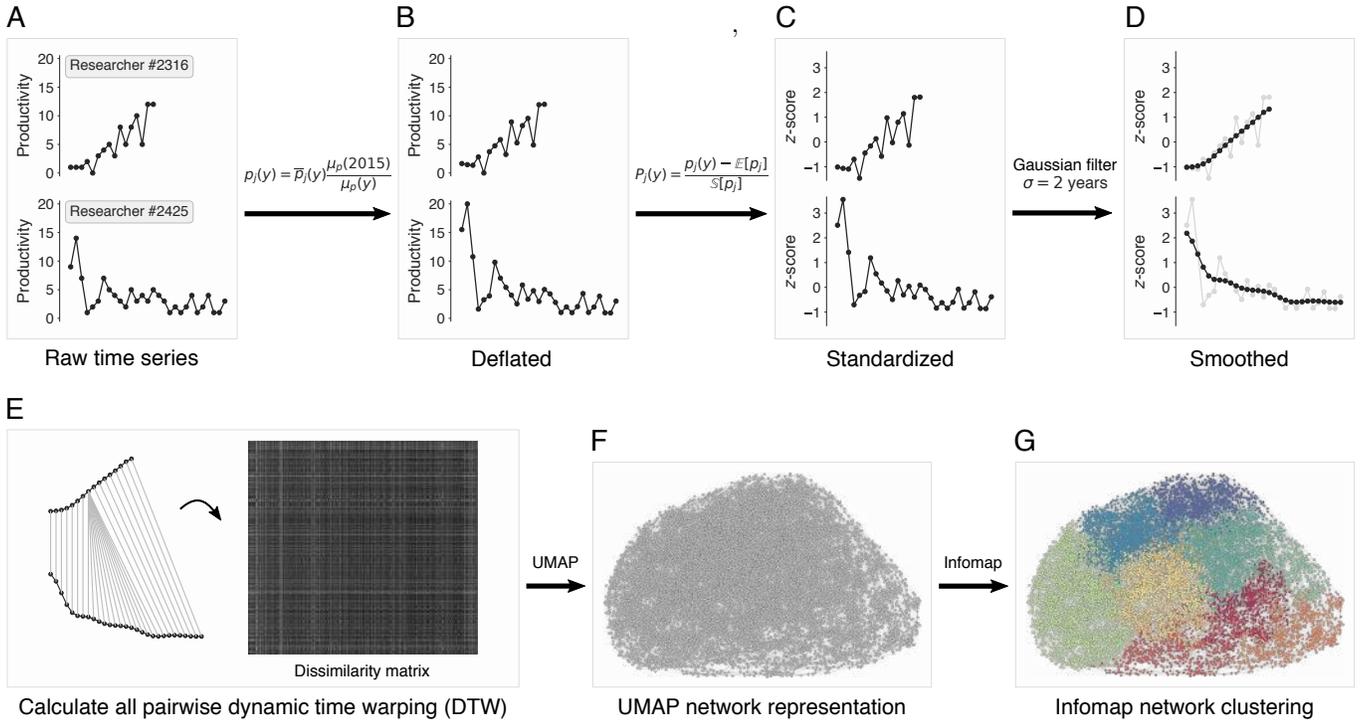

Figure S4. Description of the clustering procedure. (A) Raw productivity time series of two researchers selected as illustrative examples. (B) Deflated productivity time series. The raw time series are deflated to account for the discipline-specific inflation in productivity. The deflated productivity is obtained by dividing the raw productivity values by the corresponding average values of productivity of the researcher's discipline and multiplying the result by the average productivity of the discipline in 2015. (C) Standardized productivity time series. The deflated time series are standardized to make trajectories comparable in scale among researchers. The standardized version of the time series is obtained by subtracting the researcher's average productivity from his/her deflated productivity and dividing the result by the researcher's standard deviation of the deflated productivity. (D) Smoothed productivity time series. The time series are further smoothed to account for the noisy nature of scientific publishing. The smoothed productivity is obtained by convolving the $z$-score productivity series with a Gaussian kernel with a standard deviation of 2 years. The panel displays the time series before (gray markers) and after (black markers) the smoothing process for comparison. (E) Pairwise dissimilarity measure among researcher's productivity curves. This dissimilarity is estimated by calculating the dynamic time warping (DTW) for all pairs of productivity trajectories. The DTW allows for optimal alignment of sequences by creating a non-linear warping path between them and matching sequences with similar patterns that can be shifted in time. The left panel shows the optimal alignment of the two illustrative examples, and the right panel displays the dissimilarity matrix obtained after calculating all pairwise dissimilarities. (F) Network representation of the dissimilarity matrix. The dissimilarity matrix is fed into the uniform manifold approximation and projection (UMAP) algorithm to obtain a network representation of the productivity trajectories. Nodes represent researchers' productivity trajectories and weighted edges indicate the similarity between trajectories. The UMAP is a dimensionality reduction technique that first creates a fuzzy simplicial complex (represented as a weighted graph) and then projects the data into a lower dimensional space using a force-directed graph layout algorithm. Only the weighted graph obtained in the first step of the UMAP algorithm is used, and the lower-dimensional representation of trajectories is discarded. (G) Clusters of productivity patterns. The best partition is obtained by applying the Infomap clustering algorithm to the UMAP network representation. Infomap is a state-of-the-art network clustering technique that uses random walks as a proxy for information flow over the network, defining clusters as regions where the random walker is more likely to spend time in the network. To account for the stochastic nature of this algorithm, one thousand realizations are performed with different random seeds, and the best partition is selected as the iteration with the maximum silhouette score of the modal number of clusters. Each color represents one of the ten clusters of productivity patterns.

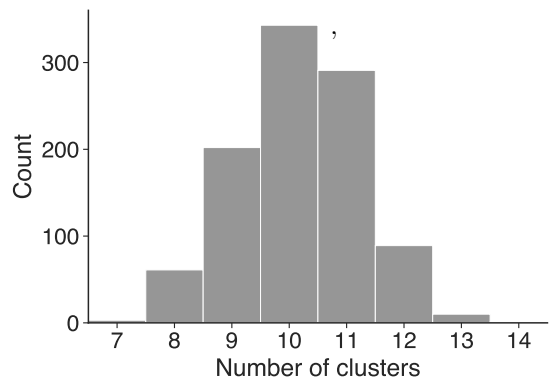

Figure S5. Histogram of the number of network communities detected in one thousand realizations of the Infomap algorithm. The number of communities varies between 7 and 14, with the majority (85%) of realizations resulting in 9 to 11 communities. The modal number of partitions, occurring in 34% of realizations, is 10.

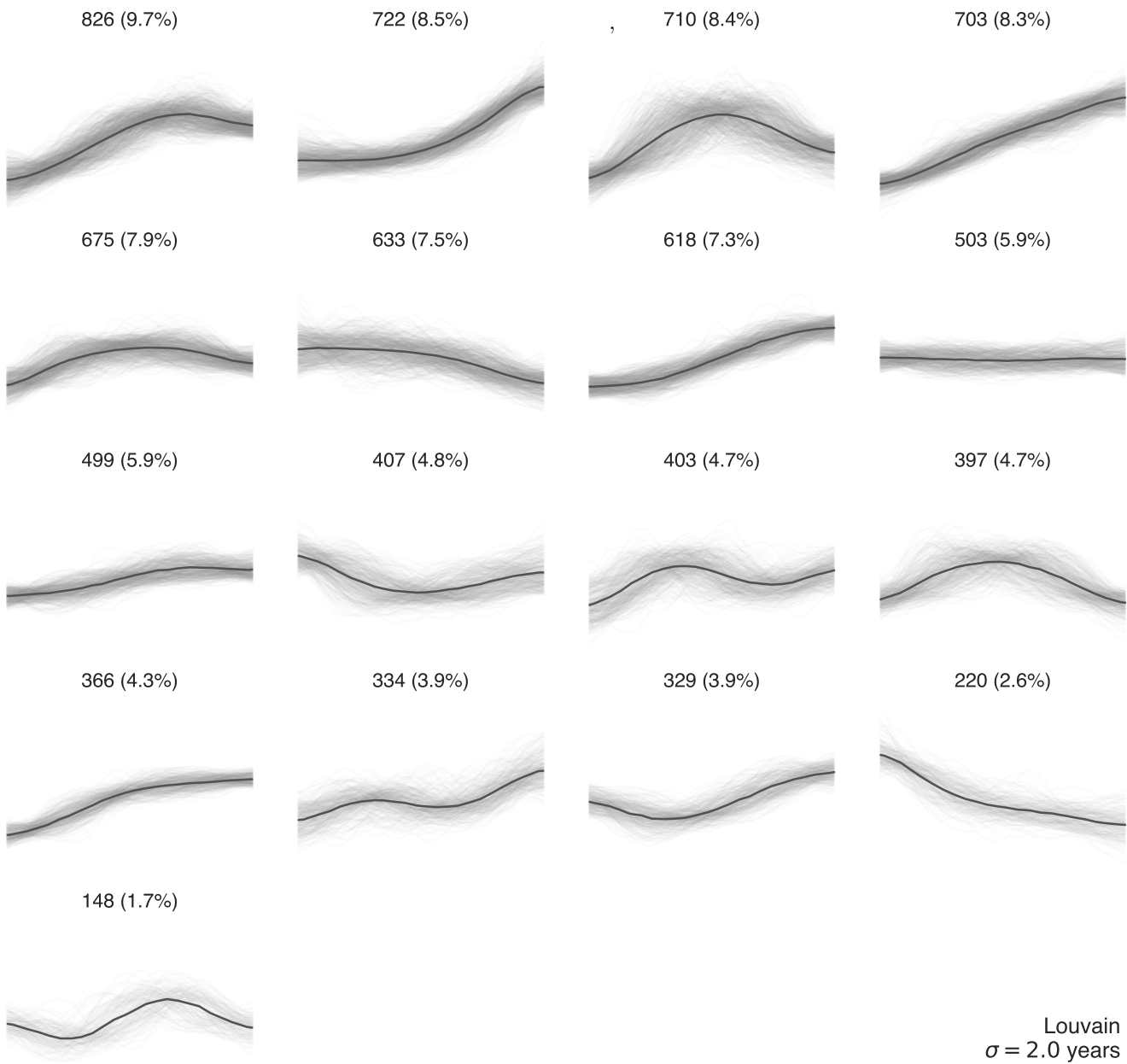

Figure S6. Clustering of researcher productivity curves using the Louvain community detection algorithm. The panels display the productivity curves of researchers in each identified community, with the black curves representing the average behavior of each cluster. The lengths of researchers' careers in each group are scaled to the unit interval, and the numbers and fractions of researchers in each group are shown within each panel. The clustering patterns obtained using the Louvain method are similar to those obtained using the Infomap algorithm.

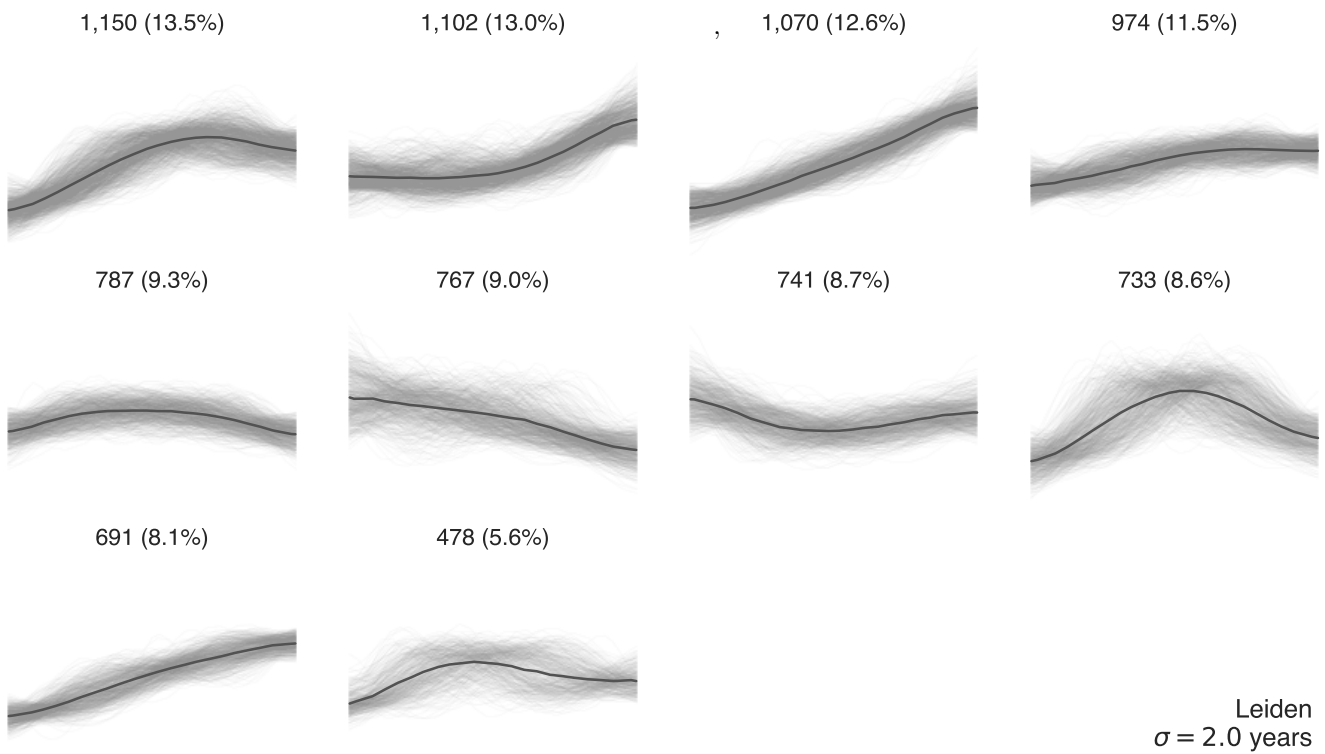

1,150 (13.5%)    1,102 (13.0%)    1,070 (12.6%)    974 (11.5%)

787 (9.3%)    767 (9.0%)    741 (8.7%)    733 (8.6%)

691 (8.1%)    478 (5.6%)

Leiden
$\sigma = 2.0$ years

Figure S7. Clustering of researcher productivity curves using the Leiden community detection algorithm. The panels display the productivity curves of researchers in each identified community, with the black curves representing the average behavior of each cluster. The lengths of researchers' careers in each group are scaled to the unit interval, and the numbers and fractions of researchers in each group are shown within each panel. The clustering patterns obtained using the Leiden method are similar to those obtained using the Louvain and Infomap algorithms.

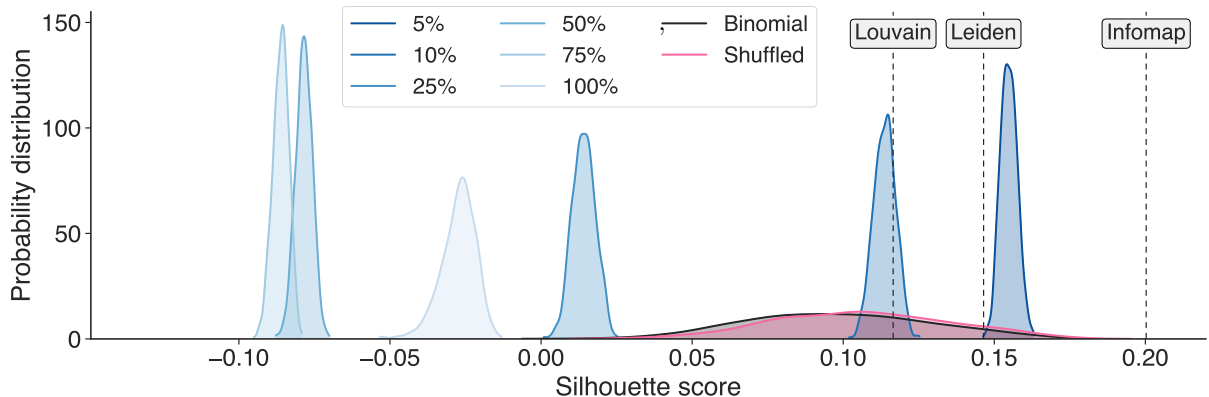

Figure S8. Assessing the consistency and significance of the clustering of productivity trajectories using the silhouette coefficient. The silhouette coefficient measures how similar each productivity series is to its own cluster when compared to other clusters. It is calculated as the normalized average difference between the cohesion (the average intracluster distance) and separation (the average nearest-cluster distance) for each series. The coefficient ranges from -1 to 1, with higher values indicating a better clustering configuration. The vertical dashed lines indicate the silhouette score for the best Infomap partition, as well as for the partitions obtained using the Louvain and Leiden algorithms. We observe that Infomap yields the highest silhouette score. The blue-shaded curves show the probability distribution of silhouette coefficients obtained after shuffling a given fraction (indicated in the legend) of the cluster labels of the Infomap partition. We observe that the silhouette decreases as we increase the fraction of shuffled labels and becomes negative before approaching zero, indicating that in the partitions obtained with high fractions of shuffled labels, the time series are closer to neighboring clusters than to their own cluster, further reinforcing the significance of the partition obtained by Infomap. To verify the significance of the clusters obtained through our procedure, we compare the silhouette score calculated from the best Infomap partition with those obtained from two null models. The first null model generates artificial trajectories using a binomial distribution with parameters set to match the average productivity of our data (4.37 papers/year). The second null model corresponds to synthetic trajectories generated by randomly shuffling the productivity trajectories of each researcher in the dataset. For each null model, we create one thousand replicas with the same number of trajectories and career size distribution of our dataset. For each replica, we apply the same procedures used to cluster the actual productivity trajectories and calculate the silhouette coefficient of the resulting partition. The black and pink curves represent the probability distributions of the silhouette scores calculated from the two null models (as indicated in the legend). The average silhouette obtained from the null models is significantly lower than the value obtained from the data, and no realization of either model yields silhouette scores higher than the one obtained from the data. These results demonstrate the significance of the clusters obtained from the actual trajectories and show that they are not algorithmic artifacts.

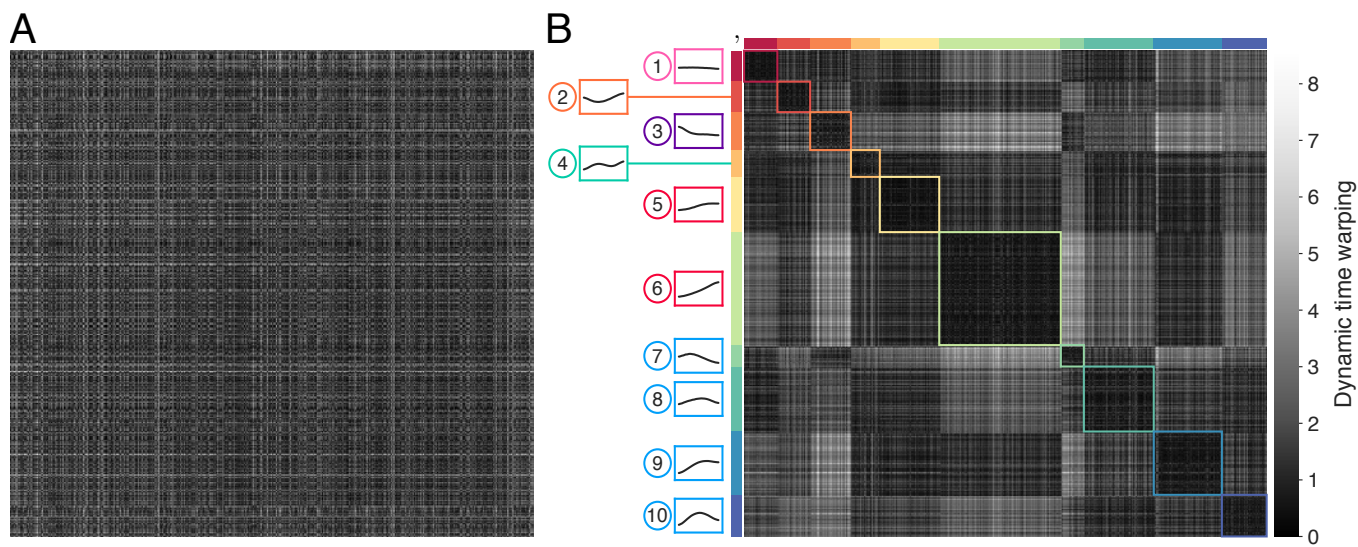

Figure S9. Dynamic time warping (DTW) dissimilarity measure among all pairs of researchers' productivity trajectories. (A) Matrix plot of the DTW dissimilarity matrix without grouping researchers using the community structure of the UMAP network. (B) Matrix plot of the DTW dissimilarity matrix after grouping researchers using the community structure of the UMAP network. The colors surrounding the matrix refer to each of the ten communities identified by Infomap. Similarly, the colored squares within the matrix representation indicate each of these groups of researchers. We observe that the communities of the UMAP network yield a block diagonal form in the dissimilarity matrix, reflecting both local and global structures of the similarities among researchers' trajectories.

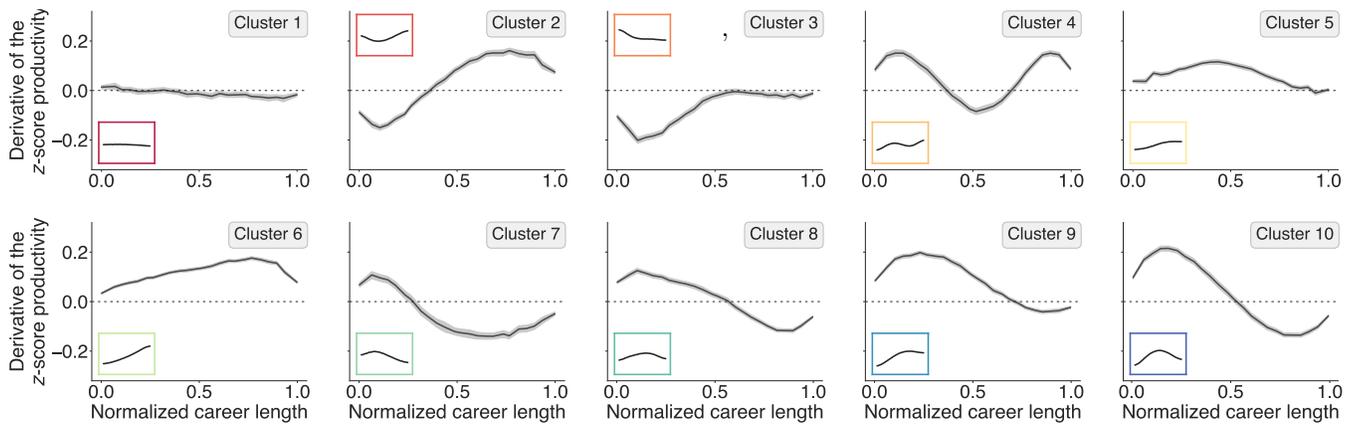

Figure S10. Derivatives of the productivity trajectories in each cluster obtained from the best Infomap partition (Figure 1 in the main text). The curves in each panel show the moving average of the differentiated productivity trajectories, with the shaded regions representing 95% confidence intervals. The lengths of researchers' careers in each group are scaled to the unit interval before estimating the averages. Positive values indicate increasing productivity rates, negative values indicate decreasing productivity rates, and values close to zero indicate constant productivity in consecutive career years. Based on these curves and the average behavior of productivity in each cluster, we have defined six categories of productivity narratives: constant (cluster 1), u-shaped (cluster 2), decreasing (cluster 3), periodic-like (cluster 4), increasing (clusters 5 and 6), and canonical-like (clusters 7 to 10).

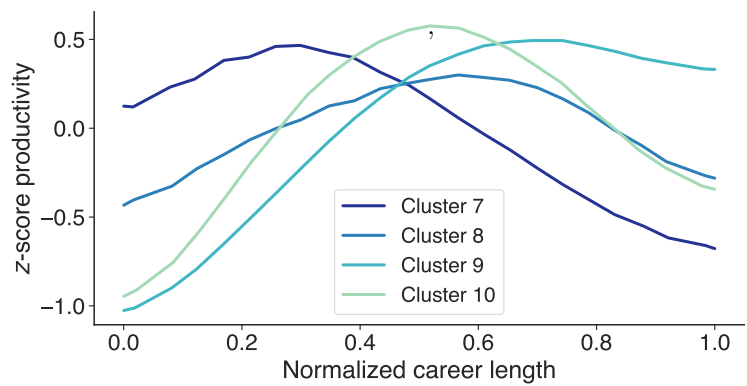

Figure S11. Differences among the average productivity behavior within the canonical-like clusters. The colored curves represent the average productivity trajectories for each of the four clusters that compose the canonical-like category (clusters 7 to 10, as indicated in the legend). The lengths of researchers' careers in each group are scaled to the unit interval. These clusters are distinguished by distinct peak positions and amplitudes, as outlined in Table S1.

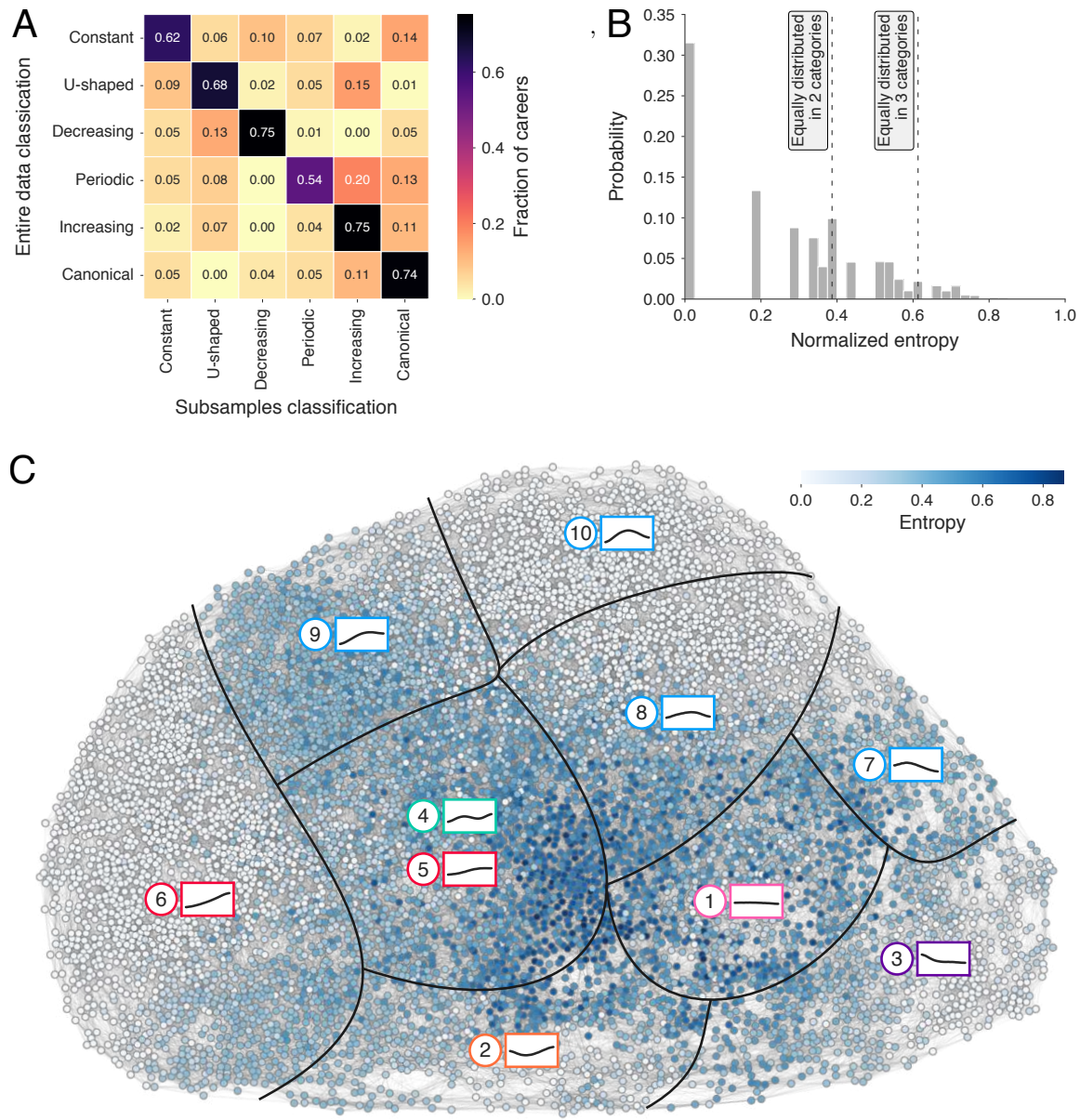

Figure S12. Validation of the robustness of the six categories of productivity patterns by subsampling the data into three equal-sized parts. The validation consists in conducting ten realizations of the clustering procedure on the subsamples obtained by dividing the complete data into three equal-sized parts. We verify that the obtained clusters can be classified into the same six categories of productivity trajectories found for the entire dataset. Researchers are then assigned to these categories and the robustness of the classification can be assessed by comparing the labels obtained from the subsamples with those of the complete dataset. (A) Average confusion matrix associated with the entire data classification (rows) and the subsamples classification (columns) calculated using the thirty subsamples. The average accuracy of the subsample classification (73%) is significantly higher than the ones of dummy classifiers using mode (39%), stratified (29%), and uniform (16%) strategies. The matrix exhibits a mostly diagonal pattern, with differences primarily occurring when periodic trajectories are labeled as increasing or canonical curves. (B) Histogram of the normalized entropy associated with the assignment probabilities of each pattern for every researcher across the ten realizations. To calculate these entropy values, we first estimate the fractions $[(p_1, p_2, \ldots, p_6)]$ that each productivity category (constant, u-shaped, decreasing, periodic, increasing, and canonical) is attributed to each researcher across the ten realizations of the subsampling strategy. Then, the normalized entropy of each researcher is calculated using the standard Shannon's entropy formula ($h = -\frac{1}{\log 6} \sum_{i=1}^{6} p_i \log p_i$). Horizontal dashed lines indicate the values of the normalized entropy for the cases in which labels are equally distributed in two or three categories for a researcher. Approximately 80% of researchers display normalized entropy below 0.5, indicating consistent classification within the same category across realizations. Additionally, about one-third of researchers present zero entropy, that is, they are always assigned to the same category. (C) Network representation where nodes represent researchers and weights correspond to the similarity between pairs of trajectories. The black lines approximately delimit the ten clusters of productivity curves (indicated in the panel by their numbers and patterns), while the blue shades correspond to the normalized entropy values. We note that researchers displaying higher entropy are more frequently located in the frontier between two or more clusters, and in the overlapping region between the periodic (cluster 4) and increasing with declining rates (cluster 5) trajectories.

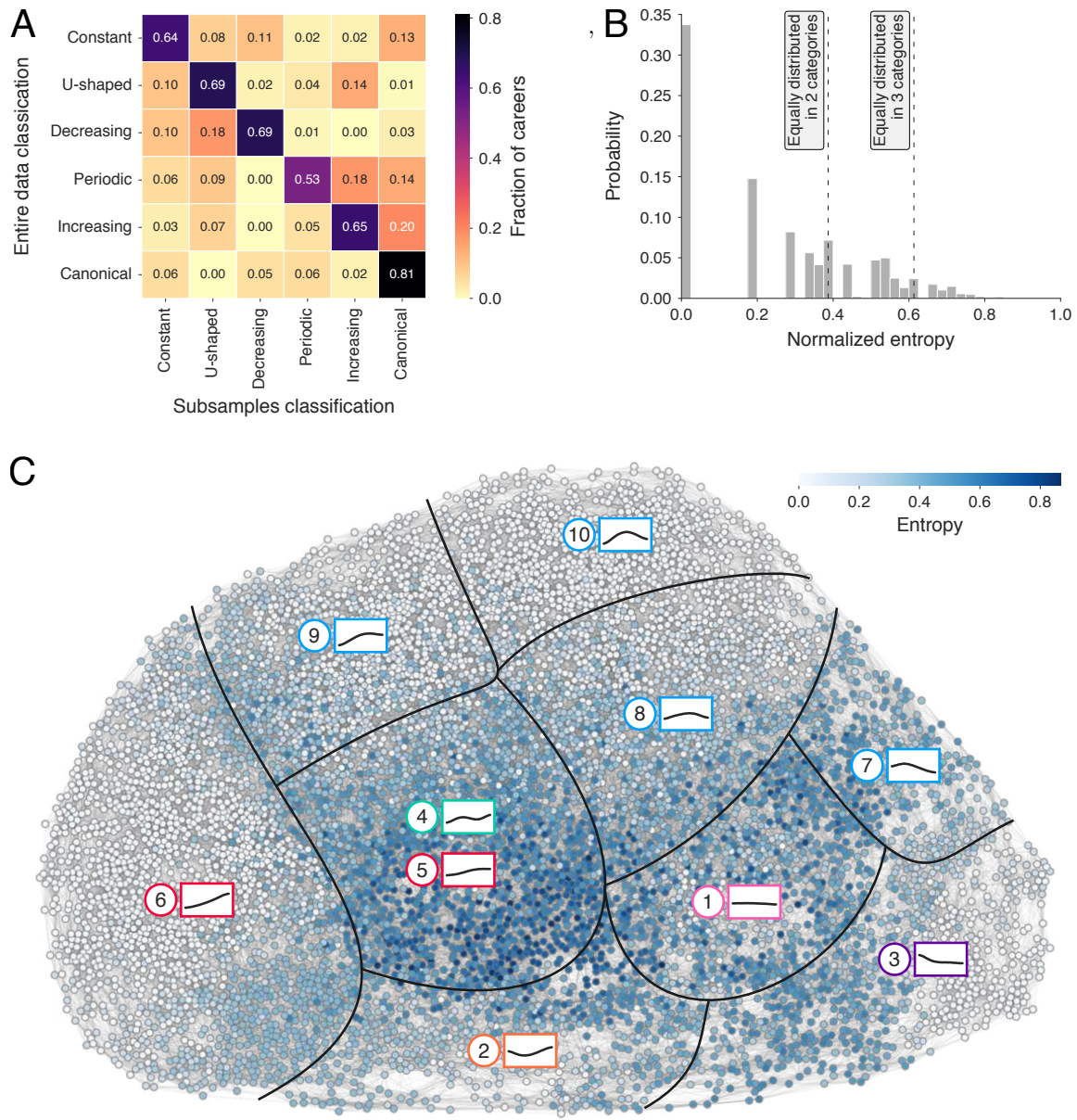

Figure S13. Validation of the robustness of the six categories of productivity patterns by subsampling the data into two halves. The validation consists in conducting ten realizations of the clustering procedure on the subsamples obtained by dividing the complete data into two halves. We verify that the obtained clusters can be classified into the same six categories of productivity trajectories found for the entire dataset. Researchers are then assigned to these categories and the robustness of the classification can be assessed by comparing the labels obtained from the subsamples with those of the complete dataset. (A) Average confusion matrix associated with the entire data classification (rows) and the subsamples classification (columns) calculated using the twenty subsamples. The average accuracy of the subsample classification (71%) is significantly higher than the ones of dummy classifiers using mode (39%), stratified (29%), and uniform (16%) strategies. The matrix exhibits a mostly diagonal pattern, with differences primarily occurring when periodic trajectories are labeled as increasing or canonical curves. (B) Histogram of the normalized entropy associated with the assignment probabilities of each pattern for every researcher across the ten realizations. To calculate these entropy values, we first estimate the fractions $[(p_1, p_2, \ldots, p_6)]$ that each productivity category (constant, u-shaped, decreasing, periodic, increasing, and canonical) is attributed to each researcher across the ten realizations of the subsampling strategy. Then, the normalized entropy of each researcher is calculated using the standard Shannon's entropy formula ($h = -\frac{1}{\log 6} \sum_{i=1}^{6} p_i \log p_i$). Horizontal dashed lines indicate the values of the normalized entropy for the cases in which labels are equally distributed in two or three categories for a researcher. Approximately 80% of researchers display normalized entropy below 0.5, indicating consistent classification within the same category across realizations. Additionally, about one-third of researchers present zero entropy, that is, they are always assigned to the same category. (C) Network representation where nodes represent researchers and weights correspond to the similarity between pairs of trajectories. The black lines approximately delimit the ten clusters of productivity curves (indicated in the panel by their numbers and patterns), while the blue shades correspond to the normalized entropy values. We note that researchers displaying higher entropy are more frequently located in the frontier between two or more clusters, and in the overlapping region between the periodic (cluster 4) and increasing with declining rates (cluster 5) trajectories.

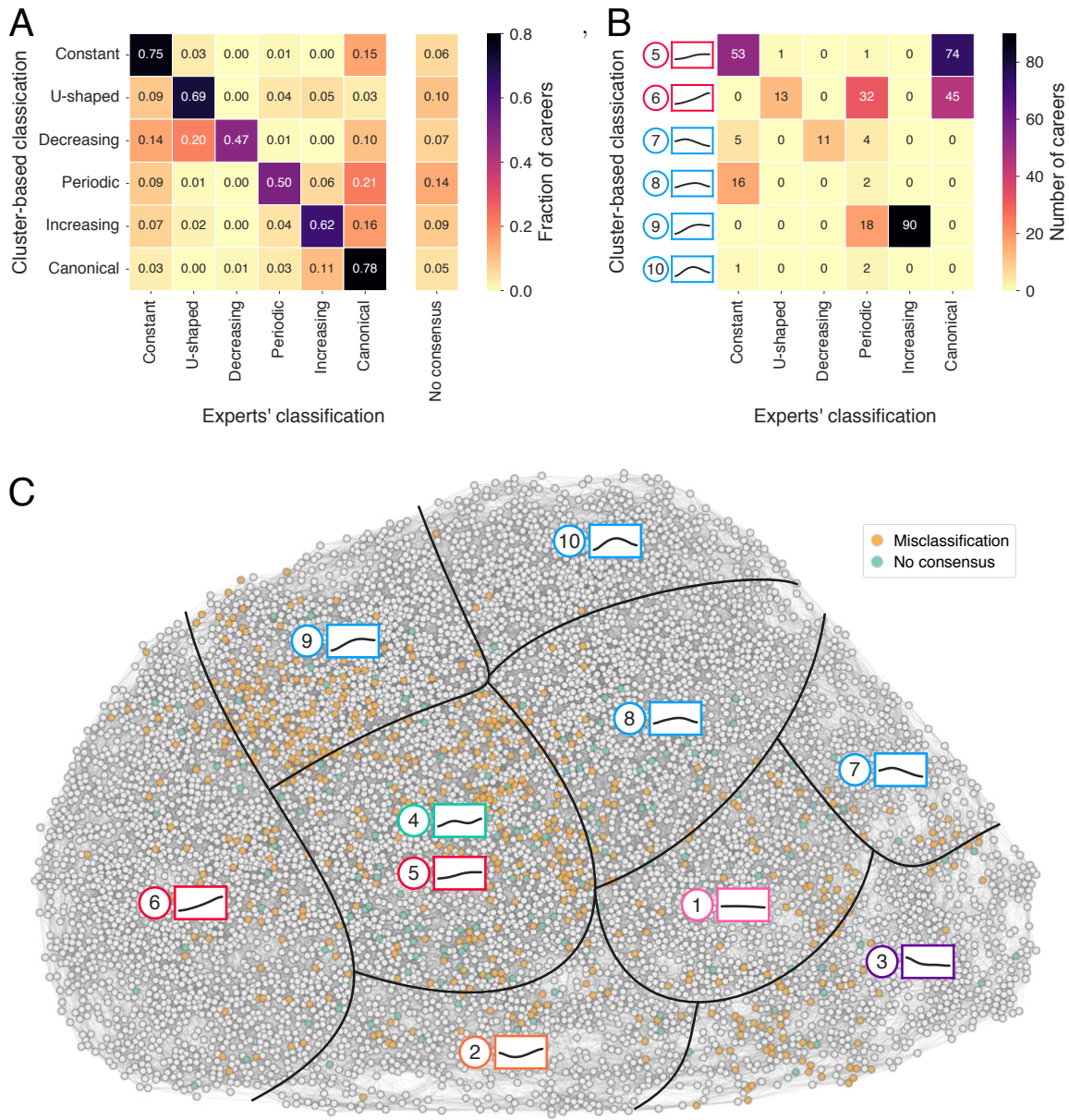

Figure S14. Human validation of the six categories of productivity patterns. A panel of two experts was recruited to classify 25% of the trajectories from our dataset, which was randomly sampled in a stratified manner. They were introduced to the average pattern of each productivity cluster (for comparison purposes) alongside an interactive application where the $z$-scores and smoothed trajectories are individually displayed. Six buttons corresponding to the productivity patterns found using the complete dataset are provided to classify trajectories alongside an additional button in case they disagree on the classification. (A) Confusion matrix associated with the cluster-based classification (rows) and the expert classification (columns). The accuracy of the experts' classification (73%) is significantly higher than the ones of dummy classifiers using mode (39%), stratified (29%), and uniform (16%) strategies. The matrix displays a mostly diagonal pattern, with differences primarily occurring when the experts classify decreasing trajectories as u-shaped curves and periodic trajectories as canonical curves. Periodic and u-shaped trajectories are proportionally the categories with the most discordance between the two experts. (B) Confusion matrix associated with the cluster-based classification (rows) and the expert classification (columns) for the clusters in the increasing and canonical categories. The increasing with declining rates (cluster 5) and late peak (cluster 9) productivity trajectories are most frequently confused with each other. (C) Network representation where nodes represent researchers and weights correspond to the similarity between pairs of trajectories. The black lines approximately delimit the ten clusters of productivity curves (indicated in the panel by their numbers and patterns). The orange markers represent misclassified trajectories (experts' classifications that do not agree with the classification obtained from our clustering procedure), while green markers indicate trajectories for which there was no consensus between the two experts. Similar to the subsampling validation analysis, both types of inconsistency are more frequently in the frontier between two or more clusters and in the overlapping region between periodic (cluster 4) and increasing with declining rates (cluster 5) patterns.

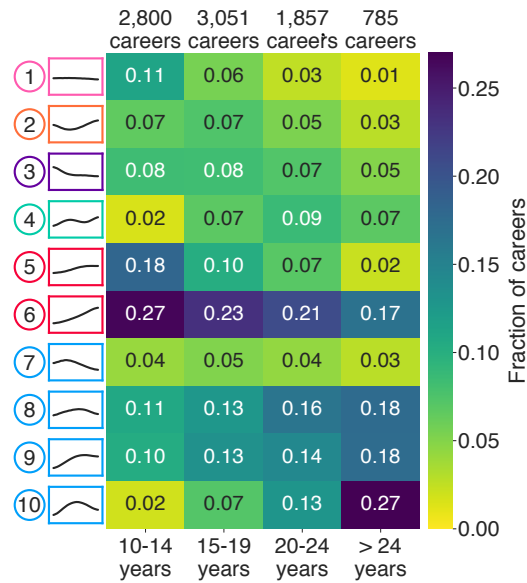

Figure S15. Career length and cohort effects on the prevalence of productivity patterns. The prevalence of productivity patterns in each of the ten clusters across four categories of career length: 10-14 years, 15-19 years, 20-24 years, and greater than 24 years. In the increasing category, we observe that the always-increasing pattern of cluster 6 is more prevalent than the increasing with declining rate pattern of cluster 5 across all categories of career length. This difference is the smallest (27% vs 18%) among the youngest cohort and the largest (17% vs 2%) among the most senior researchers. In the canonical category (clusters 7 to 10), we note that the patterns of clusters 8 and 9 are the most prevalent across all categories of career length, except among the most senior researchers, where the pattern of cluster 10 is most prevalent. The early-stage peak pattern of cluster 7 is less prevalent across all categories of career length, except among the youngest researchers, where the pattern of cluster 10 is less prevalent.

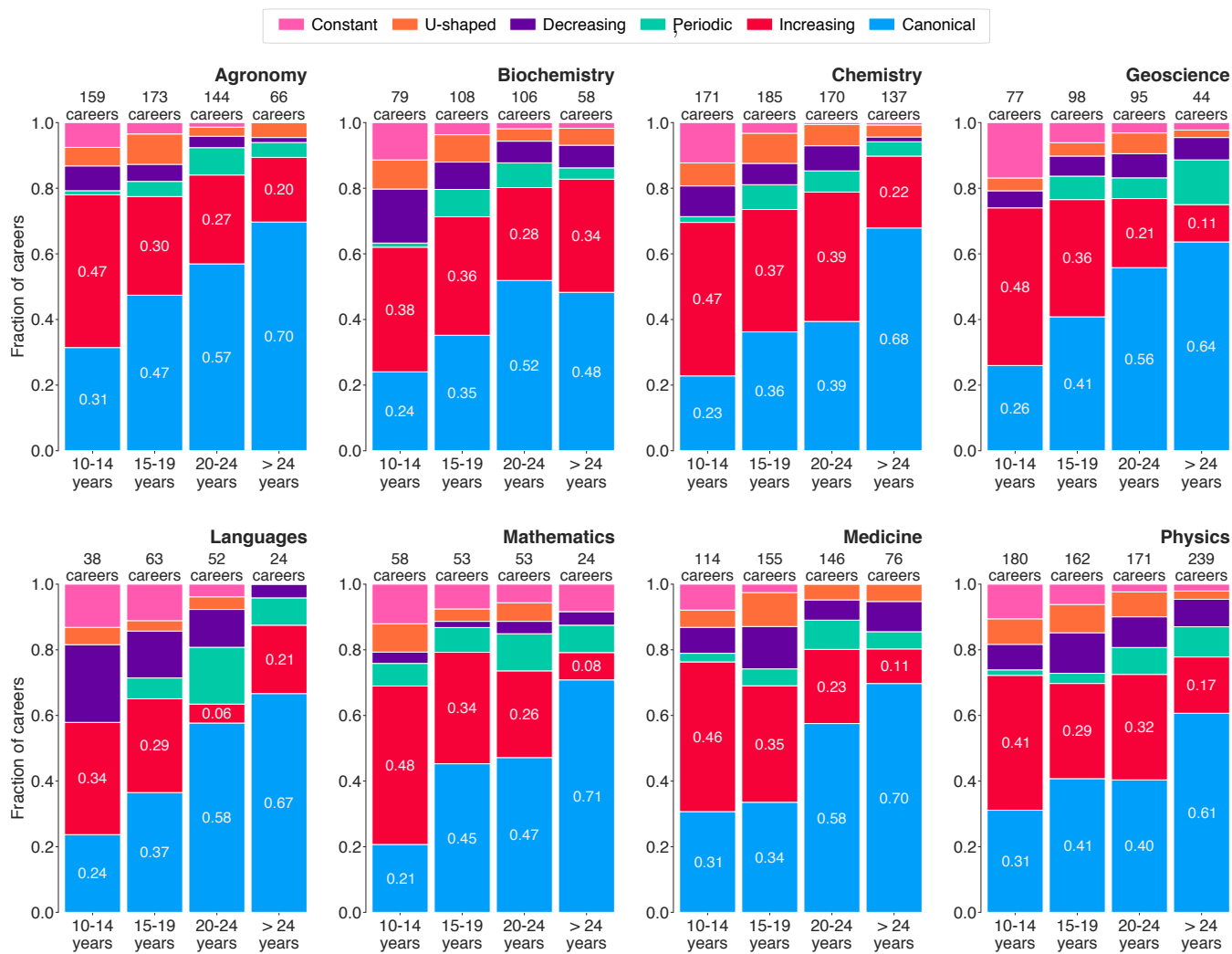

Figure S16. Prevalence of productivity patterns among different disciplines in our study. The panels show the prevalence of productivity patterns across four categories of career length (10-14, 15-19, 20-24, and greater than 24 years) for eight disciplines, each having more than twenty researchers in each length category. We note that the main patterns observed when aggregating all disciplines also emerge when considering disciplines separately. Specifically, increasing productivity curves are the dominant pattern among researchers with shorter careers, while canonical-like curves are the most prevalent pattern among senior researchers. However, there are variations in the prevalence of specific patterns among the disciplines. While 45% of all researchers among the youngest cohort show increasing productivity curves, the prevalence of this pattern is lower among researchers working on Biochemistry, Languages, and Physics (34% to 41% of researchers). For the other five disciplines (Agronomy, Chemistry, Geoscience, Mathematics, and Medicine), the prevalence of increasing patterns among the youngest researchers is only slightly higher (46% to 48% of researchers) than the fraction obtained for the aggregated case. Among the most experienced cohort, we observe that five disciplines (Agronomy, Chemistry, Languages, Mathematics, and Medicine) have a higher prevalence (from 67% to 71% of researchers) and three disciplines (Biochemistry, Geoscience, and Physics) have a lower prevalence (48% to 64% of researchers) of canonical patterns when compared with the aggregated case, in which 65% of researchers exhibit canonical-like curves. Additionally, we note that only Biochemistry shows an approximately constant fraction of increasing curves and a fraction of canonical curves that does not monotonically increase across the length categories. Except for Mathematics, which exhibits a relatively higher prevalence of constant trajectories across the length categories, all other disciplines follow the same pattern observed in the aggregated case. The prevalence of u-shaped trajectories among each discipline is similar to the pattern obtained for the aggregated case, with only Languages and Mathematics not presenting this pattern among the most experienced cohort. Decreasing trajectories display a similar decline trend across the length categories observed in the aggregated case for four disciplines (Agronomy, Biochemistry, Chemistry, and Languages), while the remaining four (Geoscience, Mathematics, Medicine, and Physics) present approximately constant fractions. Finally, the prevalence of periodic-like curves is lower among shorter careers and tends to increase across the length categories, similar to the behavior observed in the aggregated case.

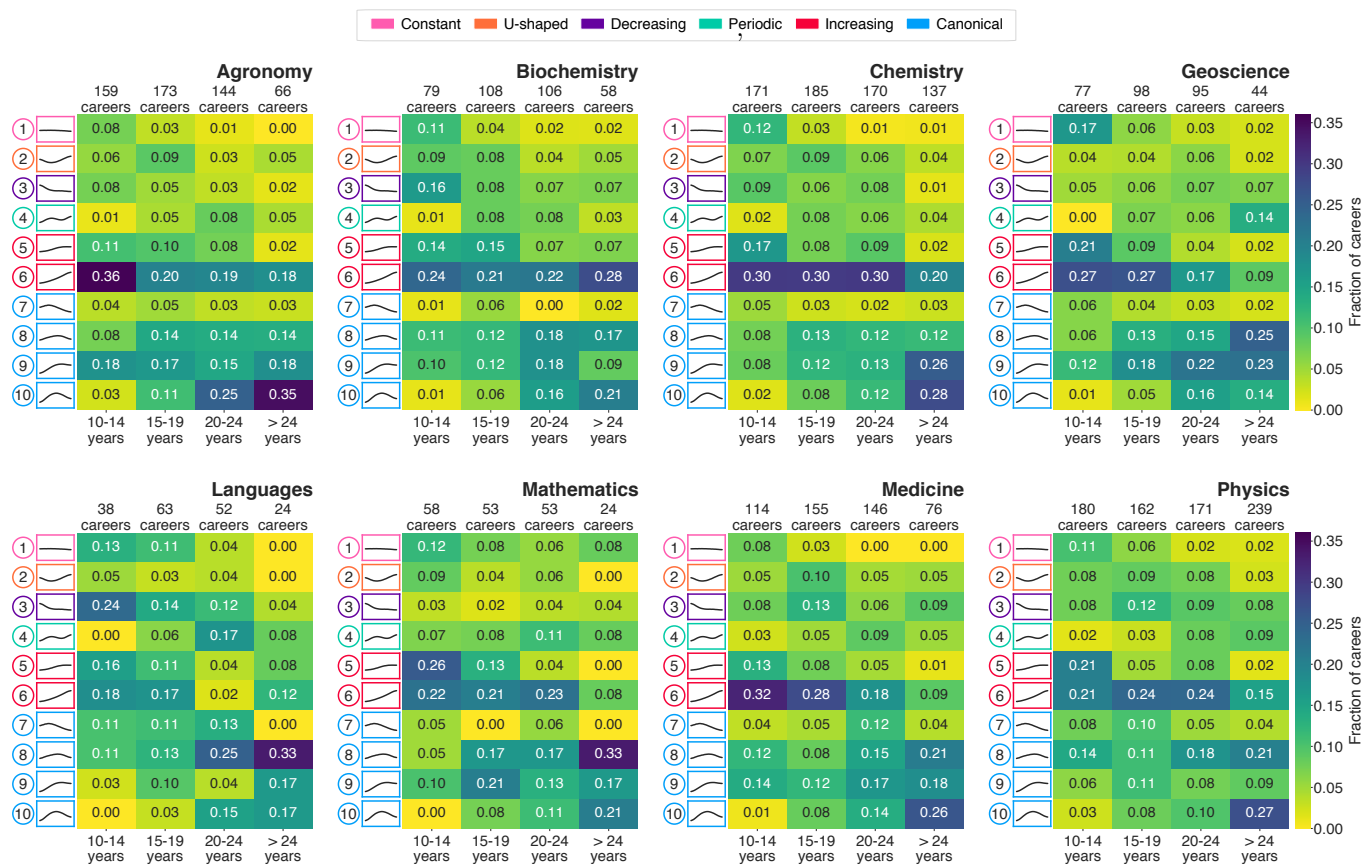

Figure S17. Prevalence of productivity patterns in each of the ten clusters among different disciplines in our study. The panels show the prevalence of each of the ten identified productivity patterns among eight disciplines, each having more than twenty researchers in four different categories of career length (10-14, 15-19, 20-24, and greater than 24 years). Similarly to the observed in the aggregated case, the always-increasing pattern of cluster 6 is more prevalent than the increasing with declining rate pattern of cluster 5 across all categories of career length and among almost all disciplines. The prevalence of these two patterns also decreases across the length categories for all disciplines but Biochemistry. In the canonical category (clusters 7 to 10), similarly to the aggregated case, we observe that clusters 8 and 9 tend to be the most prevalent patterns across all categories of career length and among most disciplines. However, while the pattern of cluster 10 is the most prevalent in the most experienced cohort among five disciplines (Agronomy, Biochemistry, Chemistry, Medicine, and Physics), the pattern of cluster 8 is most prevalent in the most experienced cohort among the remaining three disciplines (Geoscience, Languages, and Mathematics). Finally, and as in the aggregated case, cluster 7 is less prevalent among all length categories and disciplines, except for Languages which shows significantly higher fractions for the three first length categories.

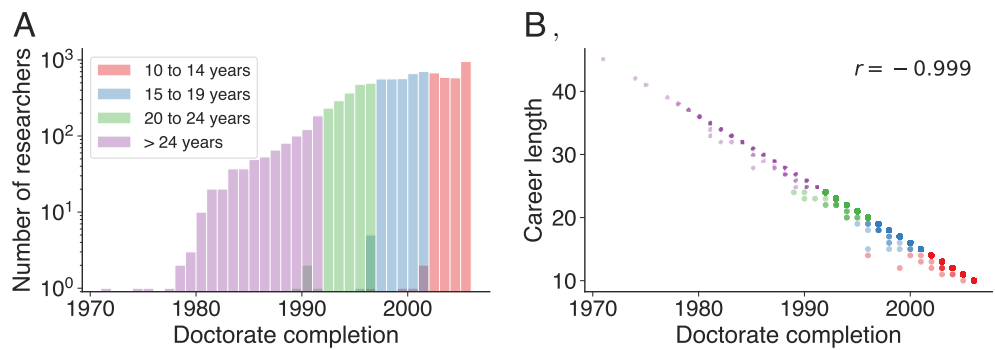

Figure S18. Career length is directly linked to the year of doctorate completion of each researcher and serves as a proxy for grouping different generations of scientists. (A) Histogram of the year of doctorate completion for all researchers in our study. (B) Direct association between career length and year of doctorate completion. These two variables are almost perfectly correlated since we assume each researcher's career starts after his/her doctorate completion. The correlation is not perfect because less than 1% of researchers had not updated their CVs in 2016 (one year before data collection).

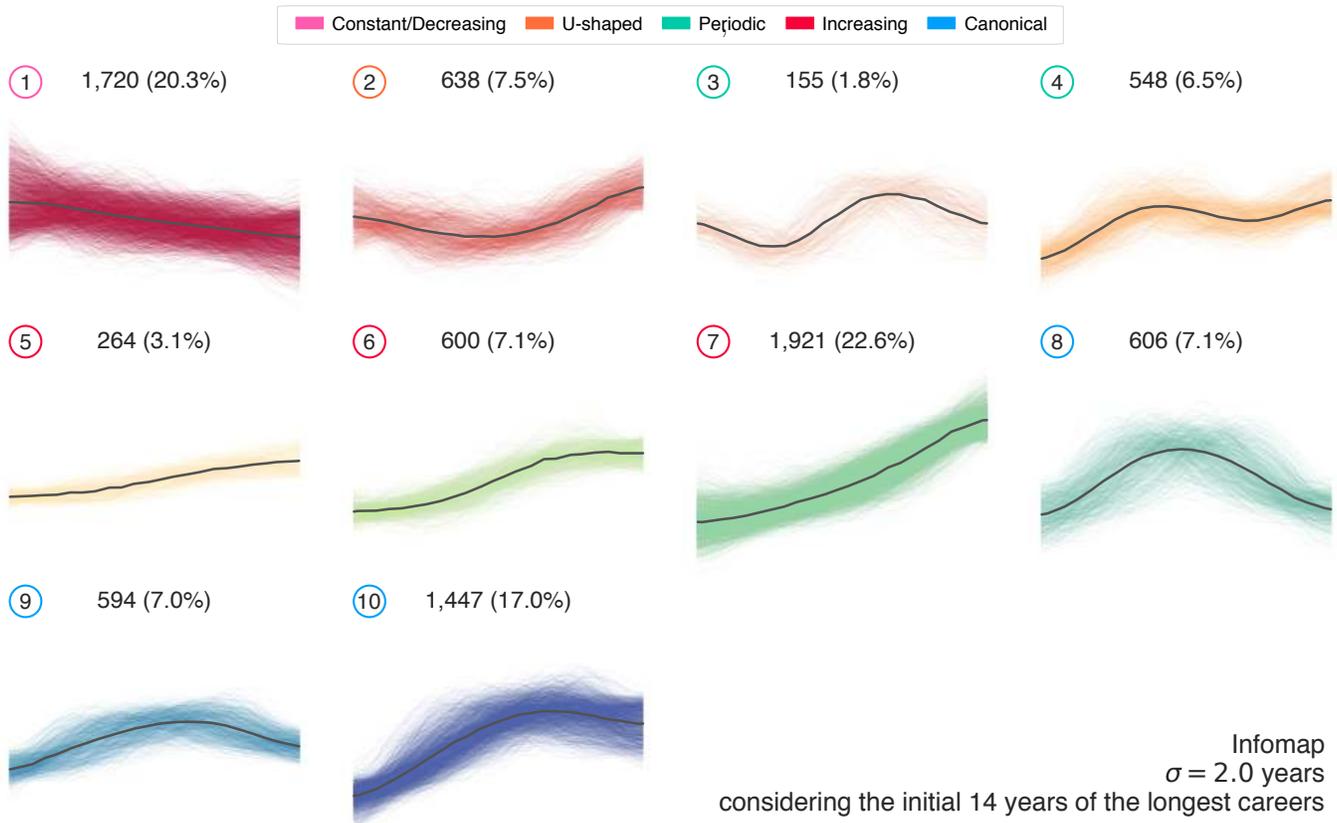

Figure S19. Clustering of researcher productivity curves obtained when applying our clustering approach to the entire dataset, but only considering the initial 14 career years of researchers with careers longer than 24 years. The panels display the productivity curves of researchers in each of the ten identified communities, with the black curves representing the average behavior of each cluster. The lengths of researchers' careers in each group are scaled to the unit interval, and the numbers and fractions of researchers in each group are shown within each panel. The ten clusters are further grouped into six categories: constant/decreasing (cluster 1), u-shaped (cluster 2), periodic-like (clusters 3 and 4), increasing (clusters 5 to 7), and canonical-like (clusters 8 to 10) curves. The clustering patterns are very similar to those obtained when considering the entire careers of senior researchers. Only the constant and decreasing patterns (clusters 1 and 2 of Figure 1 in the main text) merged into a single cluster (cluster 1) and the periodic-like curves (clusters 4 of Figure 1 in the main text) emerged as two patterns (clusters 3 and 4).

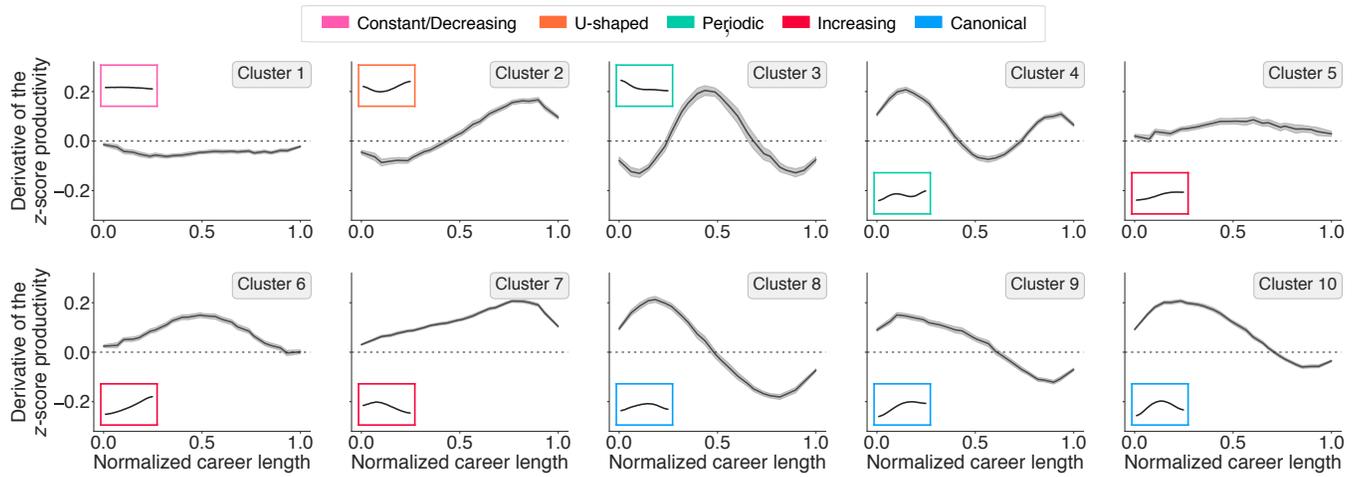

Figure S20. Derivatives of the productivity trajectories in each cluster obtained when considering the initial 14 career years of researchers with careers longer than 24 years (Fig. S19). The curves in each panel show the moving average of the differentiated productivity trajectories, with the shaded regions representing 95% confidence intervals. The lengths of researchers' careers in each group are scaled to the unit interval before estimating the averages. Positive values indicate increasing productivity rates, negative values indicate decreasing productivity rates, and values close to zero indicate constant productivity in consecutive career years. Based on these curves and the average behavior of productivity in each cluster (Fig. S19), we have defined six categories of productivity narratives: constant/decreasing (cluster 1), u-shaped (cluster 2), periodic-like (clusters 3 and 4), increasing (clusters 5 to 7), and canonical-like (clusters 8 to 10).

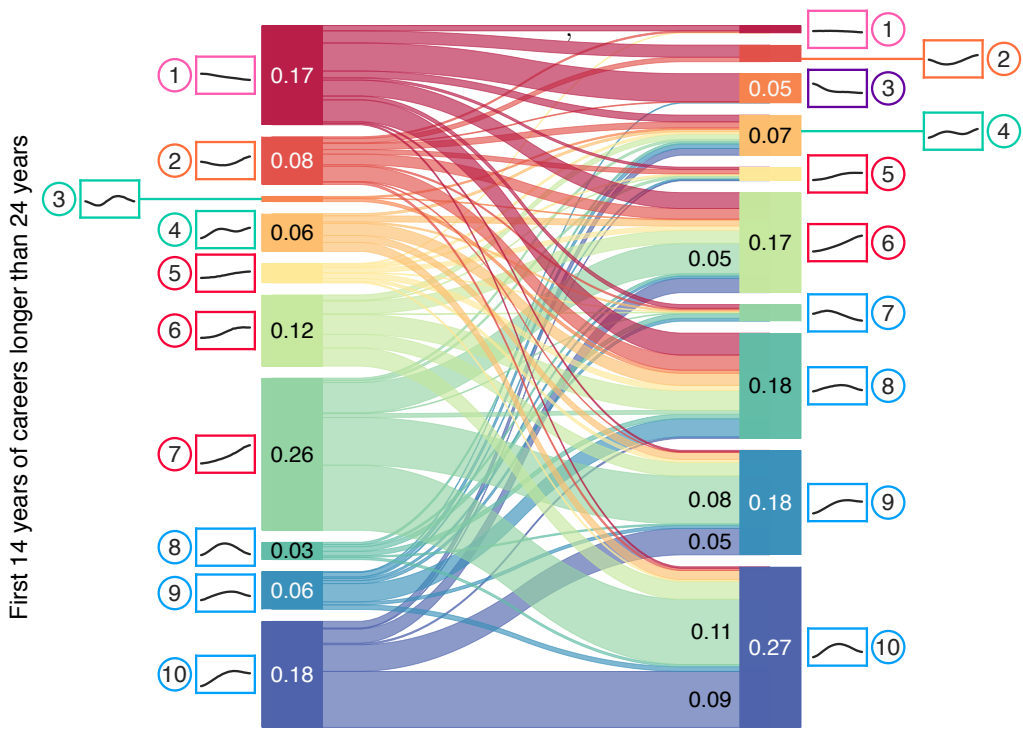

Figure S21. Comparison of the prevalence of productivity patterns in the initial career years of senior researchers with those exhibited in later career stages, when considering the individual patterns of each productivity cluster. The left bars show the fractions of productivity patterns in each cluster obtained when considering the initial 14 career years of researchers with careers longer than 24 years, and the right ones show the prevalence of patterns in each cluster when considering the full range of their careers. The connections between the left and right bars indicate the migration flow among the productivity clusters. In agreement with the results of Figure 2C of the main text, we observe that a significant part of canonical senior careers is classified as increasing curves in their beginnings, and that only a minor fraction of senior researchers who exhibit early-career increasing productivity sustain this pattern with career progression. We further observe that most transitions between increasing and canonical-like patterns occur from cluster 7 (always-increasing pattern) to clusters 9 and 10. Furthermore, the transitions between increasing patterns are the most frequent between the always-increasing patterns of clusters 7 (left side) and 6 (right side).

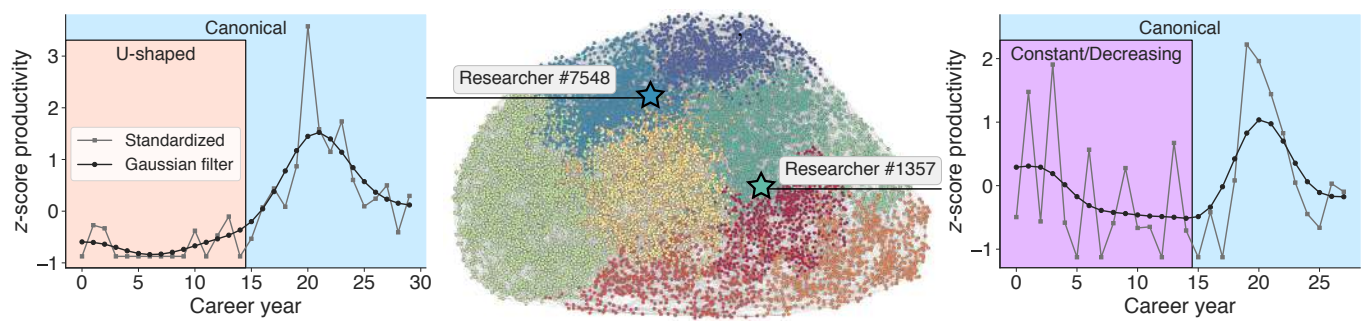

Figure S22. Examples of atypical transitions between the productivity patterns obtained when considering the initial 14 career years of researchers with careers longer than 24 years and those obtained for their entire careers. The left panel illustrates the case of a senior researcher whose productivity pattern is initially classified as u-shaped but later progresses to a canonical-like pattern. The right panel represents another atypical transition in which the career of a senior researcher is initially classified as a constant/decreasing pattern and later as a canonical-like pattern. The central panel shows the network representation of the similarities among researchers' productivity patterns, highlighting the researchers' location used to illustrate the atypical transitions. We observe that both are localized on the border between two or more communities.

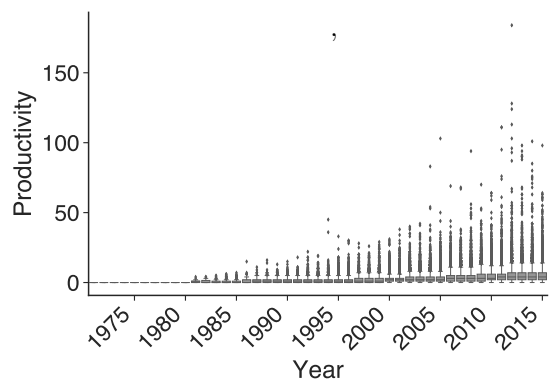

Figure S23. Box plots of raw productivity of all researchers and each year covered by our dataset. We note that extreme observations occur in almost all years. These outlier productivity values are indicated by black markers located beyond the whiskers, defined as 1.5 times the interquartile range.

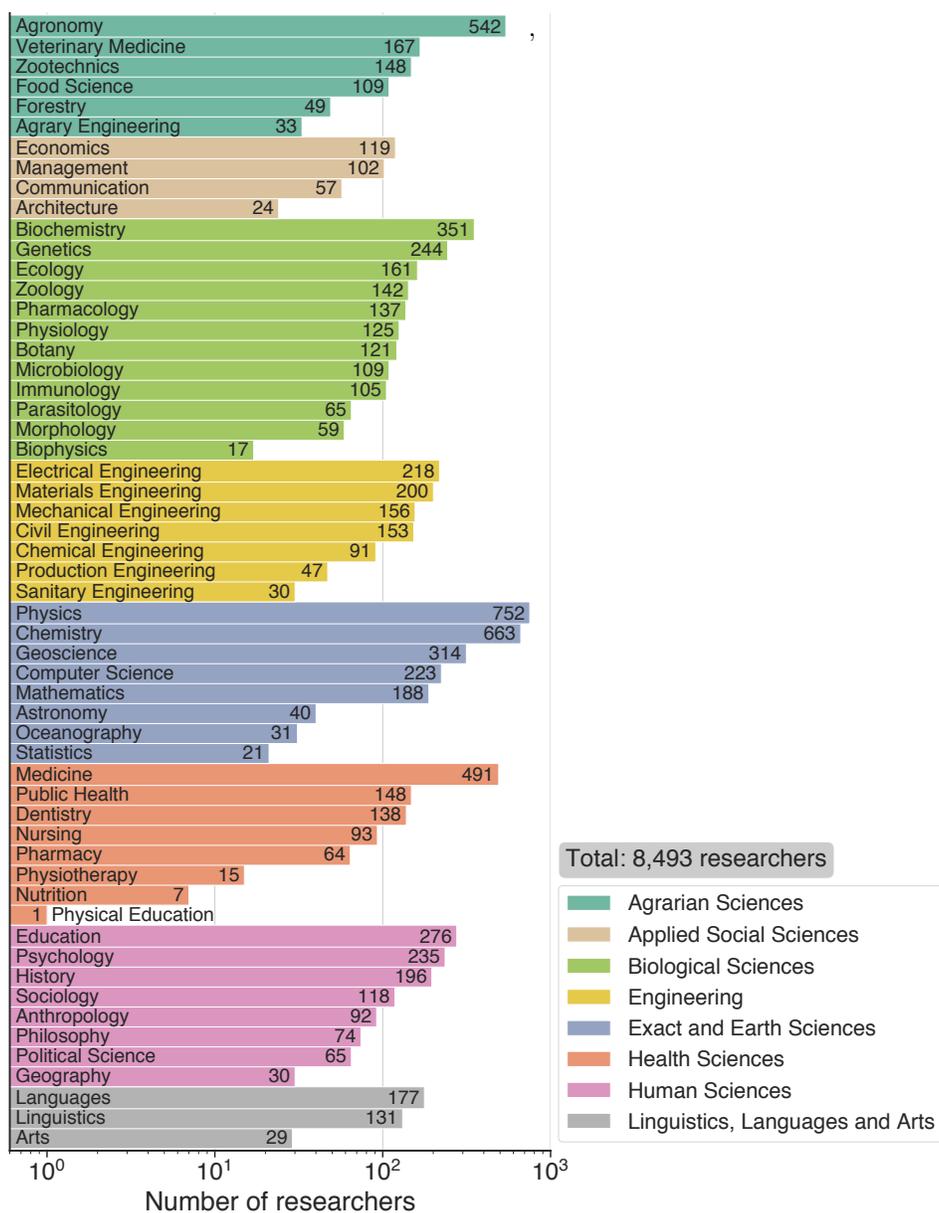

Figure S24. Number of scholars in our study and their distribution among the research disciplines. The bars show the total number of researchers for each discipline, while the color code indicates the different areas of knowledge covered by our dataset.

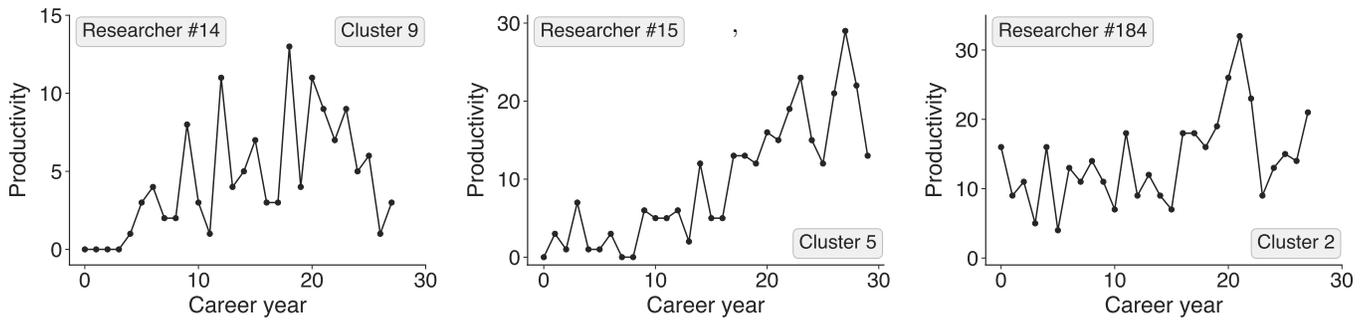

Figure S25. Illustration of the noisy nature of productivity trajectories. The panels show the raw productivity values of three selected researchers. We observe that productivity displays fluctuations which in turn reflect the complex processes involved in producing and publishing scientific papers.

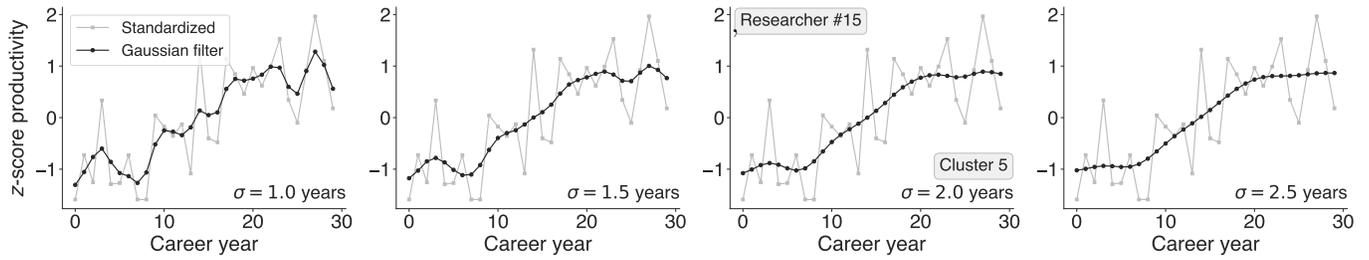

Figure S26. Illustration of the different degrees of smoothing obtained when varying the standard deviation of Gaussian kernel used to filter the $z$-score productivity series. The panel shows the $z$-score productivity series (gray markers) and the corresponding smoothed productivity series (black markers) obtained when varying the standard deviation $\sigma$ from 1.0 to 2.5 years in half-year intervals.

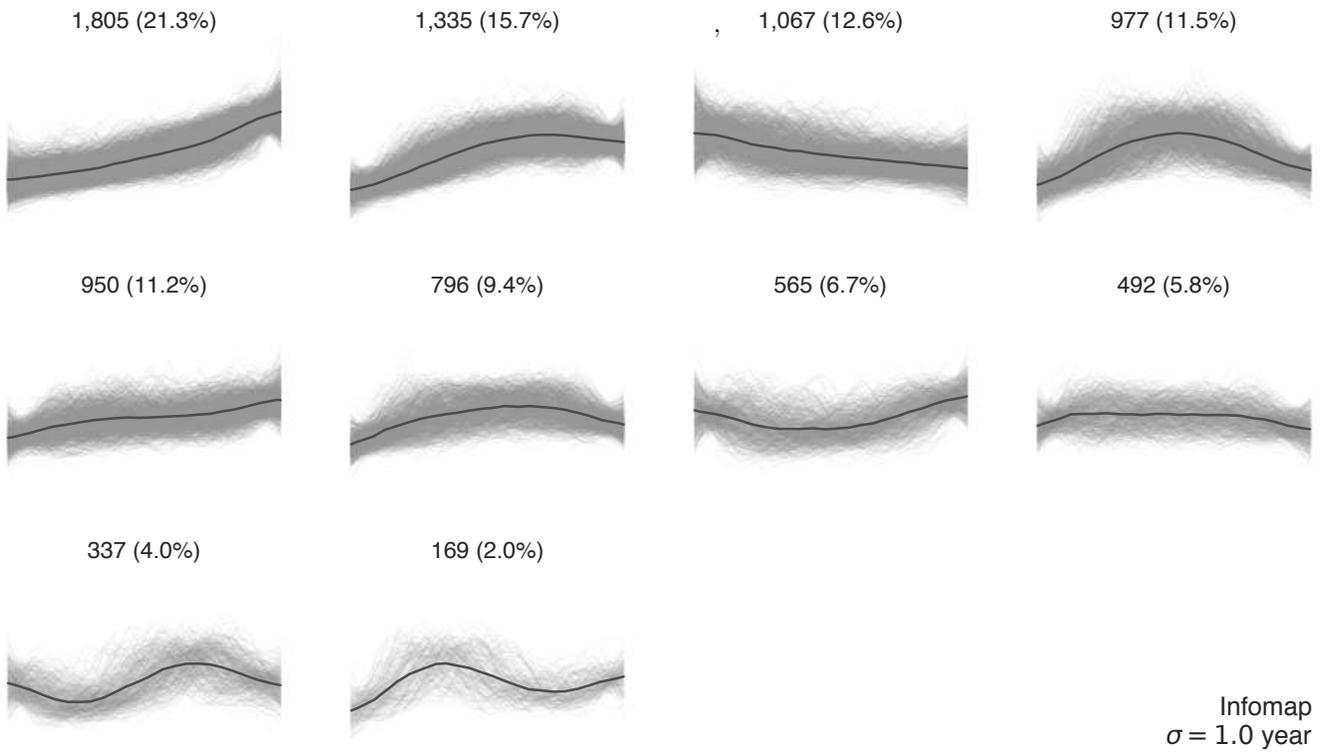

Figure S27. Clustering of researcher productivity curves using the Infomap community detection algorithm with time series smoothed using a Gaussian filter with standard deviation $\sigma = 1.0$ year. This value is smaller than the one used for the results in the main text ($\sigma = 2.0$ years). The panels display the productivity curves of researchers in each identified community, with the black curves representing the average behavior of each cluster. The lengths of researchers' careers in each group are scaled to the unit interval, and the numbers and fractions of researchers in each group are shown within each panel. The clustering patterns obtained using $\sigma = 1.0$ year is similar to those obtained for $\sigma \in \{1.5, 2.0, 2.5\}$ years.

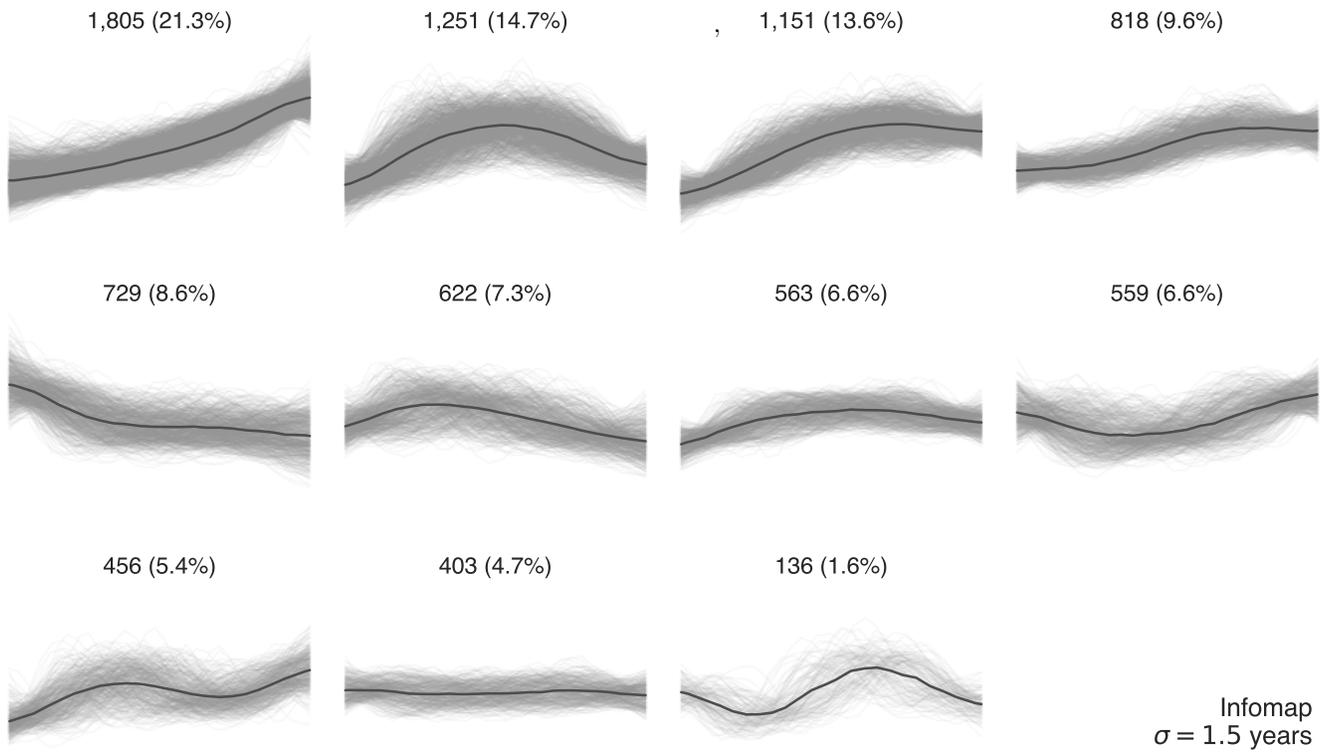

Figure S28. Clustering of researcher productivity curves using the Infomap community detection algorithm with time series smoothed using a Gaussian filter with standard deviation $\sigma = 1.5$ years. This value is smaller than the one used for the results in the main text ($\sigma = 2.0$ years). The panels display the productivity curves of researchers in each identified community, with the black curves representing the average behavior of each cluster. The lengths of researchers' careers in each group are scaled to the unit interval, and the numbers and fractions of researchers in each group are shown within each panel. The clustering patterns obtained using $\sigma = 1.5$ years is similar to those obtained for $\sigma \in \{1.0, 2.0, 2.5\}$ years.

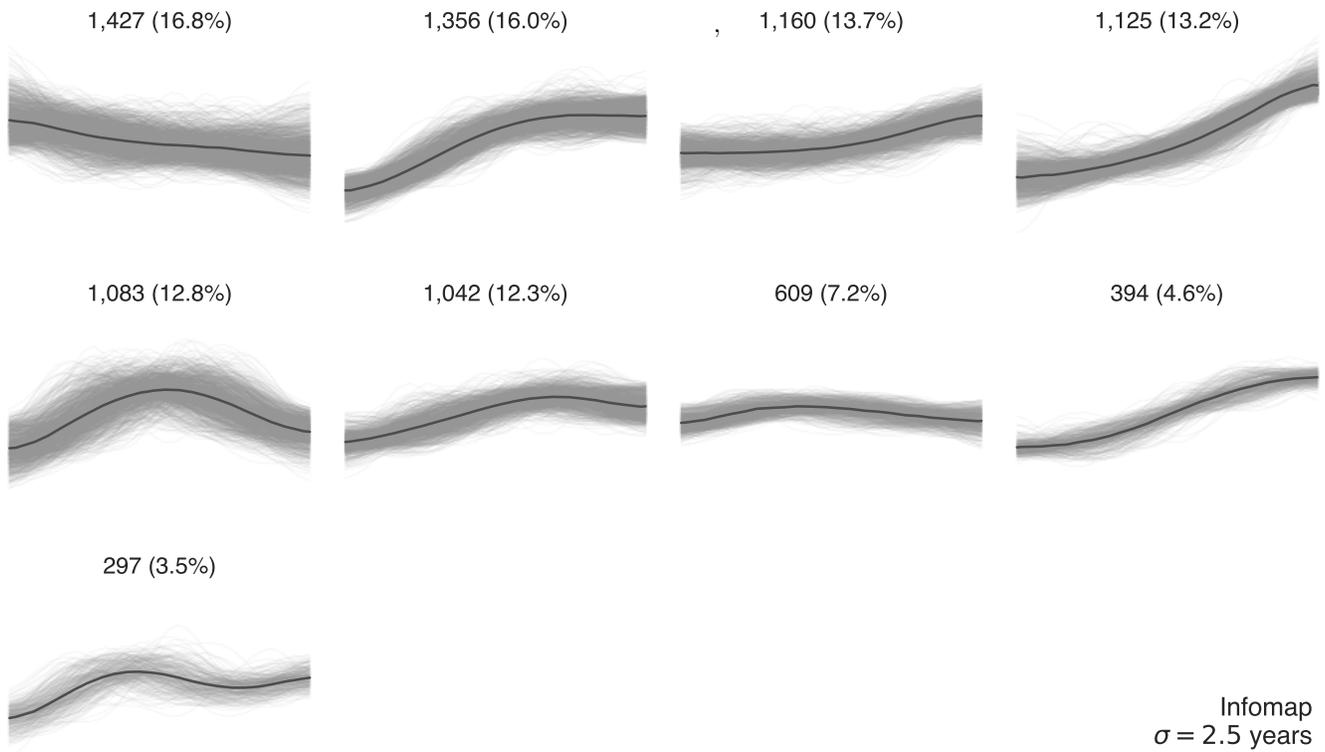

1,427 (16.8%)     1,356 (16.0%)     1,160 (13.7%)     1,125 (13.2%)

1,083 (12.8%)     1,042 (12.3%)     609 (7.2%)     394 (4.6%)

297 (3.5%)

Infomap
$\sigma = 2.5$ years

Figure S29. Clustering of researcher productivity curves using the Infomap community detection algorithm with time series smoothed using a Gaussian filter with standard deviation $\sigma = 2.5$ years. This value is larger than the one used for the results in the main text ($\sigma = 2.0$ years). The panels display the productivity curves of researchers in each identified community, with the black curves representing the average behavior of each cluster. The lengths of researchers' careers in each group are scaled to the unit interval, and the numbers and fractions of researchers in each group are shown within each panel. The clustering patterns obtained using $\sigma = 2.5$ years is similar to those obtained for $\sigma \in \{1.0, 1.5, 2.0\}$ years.

Table S1. Descriptive statistics of researchers in each productivity cluster. The amplitude column refers to the average value of the difference between the minimum and maximum productivity in standard score units. The peak position column indicates the average year of maximum productivity for all researchers in each cluster classified as a canonical-like pattern. Similarly, the normalized peak position column refers to the average position of maximum productivity after re-scaling the lengths of researchers' careers in each group to the unit interval. In these three columns, the values after the ± sign represent one standard deviation of the corresponding quantity.

| Cluster | Number of researchers | Category | Median length (years) | Amplitude (standard units) | Peak position (years) | Normalized peak position |
|---------|----------------------|----------|----------------------|----------------------------|----------------------|--------------------------|
| 1 | 542 (6.4%) | Constant | 14 | $0.58 \pm 0.19$ | - | - |
| 2 | 533 (6.3%) | U-shaped | 16 | $1.27 \pm 0.38$ | - | - |
| 3 | 663 (7.8%) | Decreasing | 16 | $1.37 \pm 0.41$ | - | - |
| 4 | 469 (5.5%) | Periodic | 19 | $1.26 \pm 0.29$ | - | - |
| 5 | 966 (11.4%) | Increasing | 14 | $1.04 \pm 0.26$ | - | - |
| 6 | 1,975 (23.3%) | Increasing | 16 | $1.91 \pm 0.32$ | - | - |
| 7 | 375 (4.4%) | Canonical | 17 | $1.37 \pm 0.29$ | $5.78 \pm 2.97$ | $0.33 \pm 0.14$ |
| 8 | 1,136 (13.4%) | Canonical | 18 | $1.16 \pm 0.30$ | $10.12 \pm 4.42$ | $0.55 \pm 0.15$ |
| 9 | 1,107 (13.0%) | Canonical | 18 | $1.72 \pm 0.28$ | $13.26 \pm 4.54$ | $0.73 \pm 0.17$ |
| 10 | 727 (8.6%) | Canonical | 21 | $1.84 \pm 0.27$ | $11.64 \pm 4.18$ | $0.54 \pm 0.13$ |



\end{document}